\documentclass[lettersize,journal]{IEEEtran}
\usepackage{amsmath,amsfonts}
\usepackage{array}
\usepackage[caption=false,font=normalsize,labelfont=sf,textfont=sf]{subfig}
\usepackage{textcomp}
\usepackage{stfloats}
\usepackage{url}
\usepackage{verbatim}
\usepackage{graphicx}
\usepackage{cite}

\usepackage{enumitem}

\usepackage{multirow}

\usepackage{graphicx} %%Grafiken in pdfLaTeX

\usepackage[labelfont=bf]{caption}

\usepackage{color}

\usepackage{tabu}

\usepackage{booktabs}

\usepackage{array}
\usepackage{url} 
\usepackage{algorithmicx}
\usepackage{tabularx,booktabs}
\usepackage{graphicx}
\makeatletter
\newif\if@restonecol
\makeatother

\usepackage[linesnumbered,ruled,vlined]{algorithm2e}%[ruled,vlined]{
\usepackage{algpseudocode}
\usepackage{amsmath}
\newcolumntype{C}[1]{>{\centering\let\newline\\\arraybackslash\hspace{0pt}}m{#1}}
\usepackage{tabularx} 
\newcolumntype{C}{>{\centering\arraybackslash}X} 
\begin{document}
\newcommand{\aliasAPP}{MB$^2$C\xspace}
\title{Multi-zone HVAC Control with Model-Based Deep Reinforcement Learning}

\author{Xianzhong~Ding,~\IEEEmembership{}
        Alberto~Cerpa,~\IEEEmembership{Member,~IEEE,}
        and~Wan~Du,~\IEEEmembership{Member,~IEEE}
        % <-this % stops a space
\thanks{A preliminary version of this work was published in the Proceedings of ACM BuildSys 2020 \cite{ding2020mb2c}.}% <-this % stops a space


}



\maketitle

\begin{abstract}
Reinforcement learning has been widely studied for controlling Heating, Ventilation, and Air conditioning (HVAC) systems.
Most of the existing works are focused on Model-Free Reinforcement Learning (MFRL), which learns an agent by extensively trial-and-error interaction with a real building.
However, one of the fundamental problems with MFRL is the very large amount of training data required to converge to acceptable performance. Although simulation models have been used to generate sufficient training data to accelerate the training process, MFRL needs a high-fidelity building model for simulation, which is also hard to calibrate.
As a result, Model-Based Reinforcement Learning (MBRL) has been used for HVAC control.  While MBRL schemes can achieve excellent sample efficiency (i.e. less training data), they often lag behind model-free approaches in terms of asymptotic control performance (i.e. high energy savings while meeting occupants' thermal comfort). 
In this paper, we conduct a set of experiments to analyze the limitations of current MBRL-based HVAC control methods, in terms of model uncertainty and controller effectiveness. Using the lessons learned, we develop \aliasAPP, a novel MBRL-based HVAC control system that can achieve high control performance with excellent sample efficiency. \aliasAPP learns the building dynamics by employing an ensemble of environment-conditioned neural networks.  It then applies a new control method, Model Predictive Path Integral (MPPI), for HVAC control. It produces candidate action sequences by using an importance sampling weighted algorithm that scales better to high state and action dimensions of multi-zone buildings. We evaluate \aliasAPP using EnergyPlus simulations in a five-zone office building.  The results show that \aliasAPP can achieve 8.23\% more energy savings compared to the state-of-the-art MBRL solution while maintaining similar thermal comfort.  \aliasAPP can reduce the training data set by an order of magnitude (10.52$\times$) while achieving comparable performance to MFRL approaches.

\end{abstract}


\section{Introduction}
People usually spend 87\% of their time inside building \cite{klepeis2001national, rajabi2022modes}. To maintain thermal comfort in buildings, a significant amount of energy is used to condition the thermal zones in the building. 
Buildings account for 40\% of energy usage in the US and 50\% of that energy goes to Heating, Ventilation, and Air Conditioning (HVAC)~\cite{energy_hvac}.
Advanced control strategies for operating HVAC systems have the potential to significantly improve the energy efficiency of our building stock and enhance comfort\cite{bengea2012model, goetzler2017energy}.

Rule-based Control (RBC) is widely used to set actuators (e.g., heating or cooling temperature, and fan speed) in HVAC systems~\cite{salpakari2016optimal}. One of the main advantages is that they are easy to understand.
However, RBC "rules" are usually set some if-then rules using many times static thresholds based on the rule-of-thumb rules and the experience of engineers and facility managers.  They have two fundamental problems: first, they do not scale well with the problem size, as the buildings become larger and more complex, rules must be added; second, they do not handle incomplete or incorrect information very well, an occurrence common in buildings in practice; and finally, they do not necessarily provide a guarantee of optimal control.

Model Predictive Control (MPC) has been widely studied to address these drawbacks by finding optimal control actions based on an analytical building model~\cite{beltran2014optimal, winkler2020office}. Normally, an optimization problem is formulated with the building model and some constraints, and analytic gradient computation is used to optimize over actions and building states simultaneously. 
However, this often requires convexification of the cost function and first or second-order approximations of building dynamics~\cite{kota1996optimal} in order to solve the optimization problem fast and to scale well. 
And also, the complexity of thermal dynamics and various influencing factors (e.g., layouts, HVAC configurations and occupancy pattern) are hard to be precisely captured in analytical energy models for heterogeneous buildings~\cite{lu2015modeling}.
As a result, the models used in current solutions are simplified to deal with the parameter-fitting data requirement and computational complexity~\cite{beltran2014optimal, winkler2020office}.
For example, Gnu-RL~\cite{chen2019gnu} adopts a differentiable MPC policy; whereas it uses a simple linear model to represent the dynamics of a water-based radiant heating system.

Reinforcement Learning (RL) has been widely studied for HVAC control~\cite{ding2019octopus,zhang2018practical, nagy2018reinforcement, ding2023exploring, park2020hvaclearn}. 
RL adapts to different environments by learning a control policy through direct interaction with the environment~\cite{sutton2018reinforcement}.
Current solutions mainly adopt Model-Free Reinforcement Learning (MFRL), which learns an optimal HVAC control policy by trial-and-error interactions with a real building. 
However, MFRL requires a large amount of interactions to converge, e.g., in our experiments, it requires 500,000 timesteps (5200 days) to achieve a high control performance.
Although a simulated building model can be used to accelerate the training process, it needs a high-fidelity model, which is hard to calibrate~\cite{zhang2018practical, ding2019octopus}. 
Recently, Model-Based Reinforcement Learning (MBRL) has been tested for HVAC control to achieve high data efficiency~\cite{zhang2019building}.
The HVAC system dynamics is first learned using a neural network based on historical HVAC data. 
Based on the learned building dynamics model, an MPC controller tries to find the optimal control action by using a Random Shooting (RS) method~\cite{zhang2019building}. 
For controlling a single-zone HVAC system, an MBRL-based approach saves approximately 10$\times$ training time of the MFRL approach, while achieving comparable performance~\cite{zhang2019building}.
However, most of the commercial buildings are multi-zone buildings~\cite{goyal2012method}.  In addition to the above scheme not being suitable for multi-zone HVAC systems, MBRL often lags behind the MFRL schemes in terms of control performance (high energy saving while meeting the thermal comfort of occupants).

To overcome these limitations, this paper presents \aliasAPP, a novel MBRL-based HVAC control approach that can achieve both the data/sample efficiency of MBRL and the control performance of MFRL. 
The design goal of \aliasAPP is to meet the thermal comfort requirements of the occupants while saving as much energy as possible.
The energy consumed by a building HVAC system and the thermal comfort of occupants are determined by a set of factors, including current state of all zones, the outdoor weather and the control actions we are about to take (e.g. temperature setpoints). 
In a multi-zone building, the control actions can be represented as a vector $A_{s}$, which is a combination of control actions for all thermal zones.
\aliasAPP finds the best $A_{s}$ from all possible action combinations $A_{all}$ for each control cycle. 
The best A$_{s}$ maintains the thermal comfort in its acceptable range for the entire control interval with the lowest energy consumption. 
\aliasAPP is mainly composed of two parts: (a) a building dynamics model, and (b) an HVAC control algorithm. 

Our building dynamics model employs an ensemble of environment-conditioned neural networks. We use a neural network model that takes the current state of the building and the action to perform as input, and outputs a prediction of the next state of the building. 
To capture model uncertainty, we design a novel weighted ensemble learning algorithm that aggregates the results of multiple building dynamics models by dynamically adjusting the weight of each model according to their accuracy. 
We also adopt an environment-conditioned neural network architecture by separating the action-depended state items (e.g., zone temperature) and the environment-related state items (e.g., outside temperature), since the latter cannot be actuated by control actions. 

Based on a learned building dynamics model, a flexible way to solve the control optimization problem is a shooting method that samples stochastic action trajectories for a number of incoming time-steps~\cite{nagabandi2018neural}. 
An action trajectory is a set of actions for incoming $H$ time-steps.
Every time, $H$ time-steps are evaluated, but only the first action will be executed at the next time-step.
For example, RS has been used in the latest MBRL-based HVAC control solution~\cite{zhang2019building}, which entails sampling candidate actions from a uniform distribution. 
However, RS is insufficient to find the best action trajectory, because randomly-shot action trajectories may not include it. 
We adopt Model Predictive Path Integral (MPPI) control method, which has shown promising performance in robotics control~\cite{williams2017information}. 
MPPI derives an optimal control action as the first action of a noise-weighted average over sampled control action trajectories by changing the initial control input and variance of the sampling distribution. 
We customize MPPI control for building HVAC control under the MBRL-based framework with the best parameter setting.

We implement \aliasAPP in Tensorflow, an open-source machine learning library in Python, with a 3-layer neural network as the building dynamics model and an MPPI-based control algorithm. 
We study the performance of \aliasAPP and compare it with benchmark methods by controlling a building of five thermal zones. 
We conduct a variety of simulations in EnergyPlus for evaluation. 
Extensive simulations reveal that \aliasAPP outperforms the latest model-based DRL method by 8.23\% in total energy consumption of the building, without scarifying thermal comfort. Compared with the model-free DRL approach, we reduce the training convergence time by 10.52$\times$, more than an order of magnitude improvement.

 \begin{figure*}[t]
		\begin{minipage}[t]{0.32\linewidth}
		\centering
		 \includegraphics[width=2.2in,height=1.35in,angle=0]{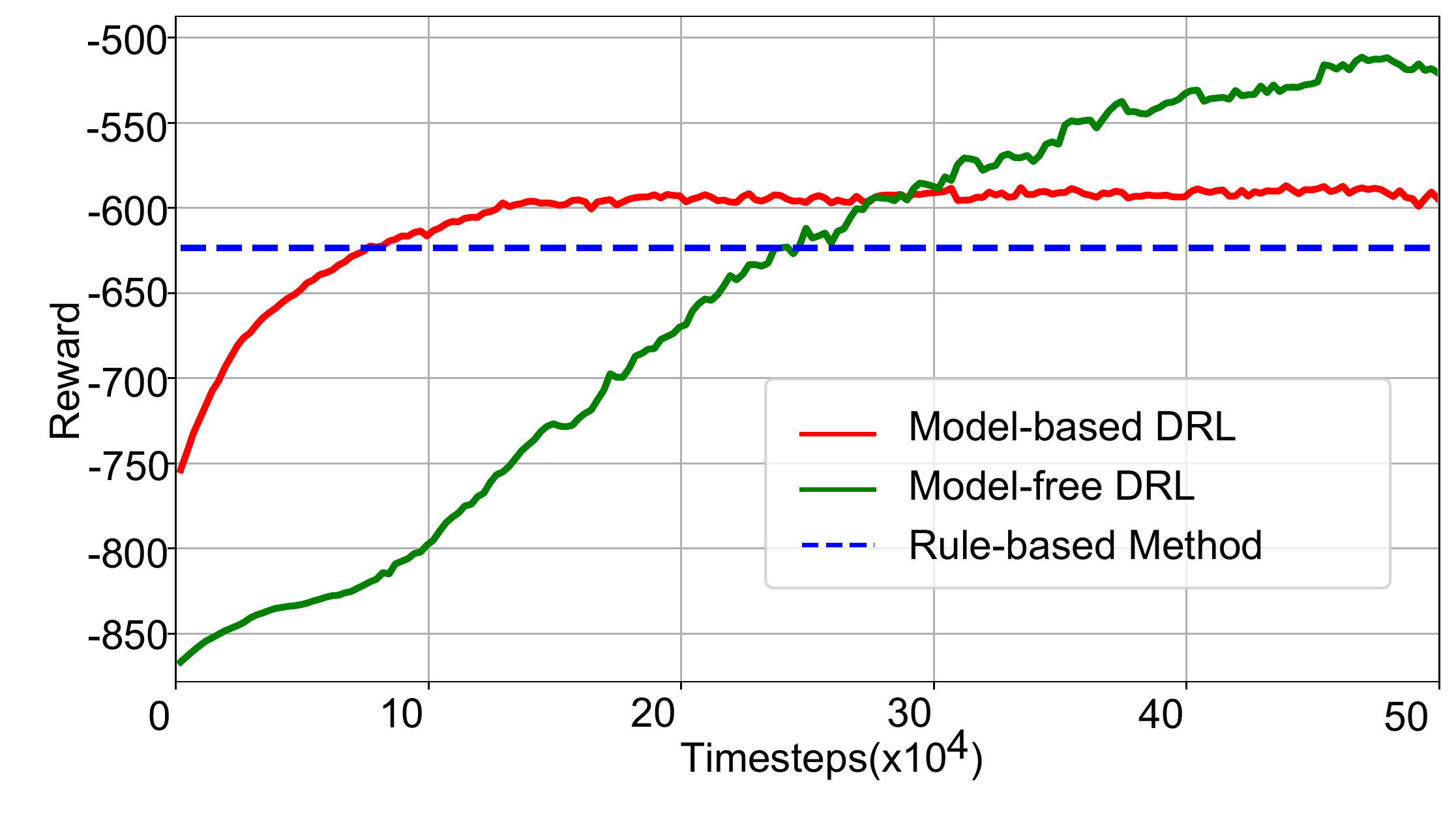}
% 		 \vspace{-0.1in}
		\caption{Convergence time and the achieved reward.}
		\label{mf_better_mb}
	\end{minipage}	
		\hspace{1ex}
	\begin{minipage}[t]{0.32\linewidth}
		\centering
		\includegraphics[width=2.2in,height=1.35in,angle=0]{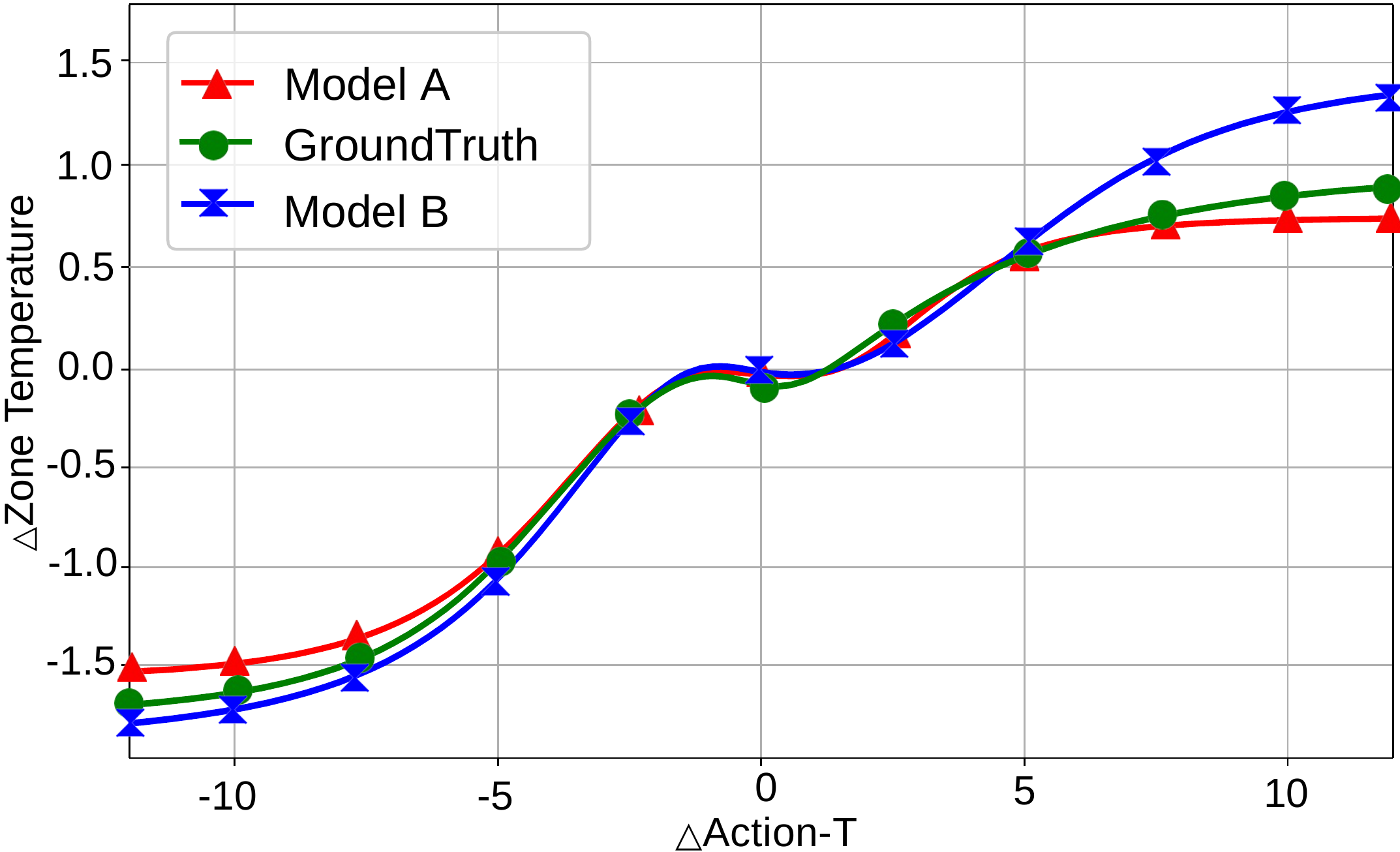}
		% \includegraphics[width=2.2in,height=1.35in,angle=0]{samples/ensemble_motivation_new.pdf}
% 		 \vspace{-0.1in}
		\caption{Uncertainty of the building dynamics model.}
		\label{aleatoric_epistemic_uncertainty}
	\end{minipage}	
	\hspace{1ex}
	\begin{minipage}[t]{0.32\linewidth}
		\centering
		 \includegraphics[width=2.2in,height=1.35in,angle=0]{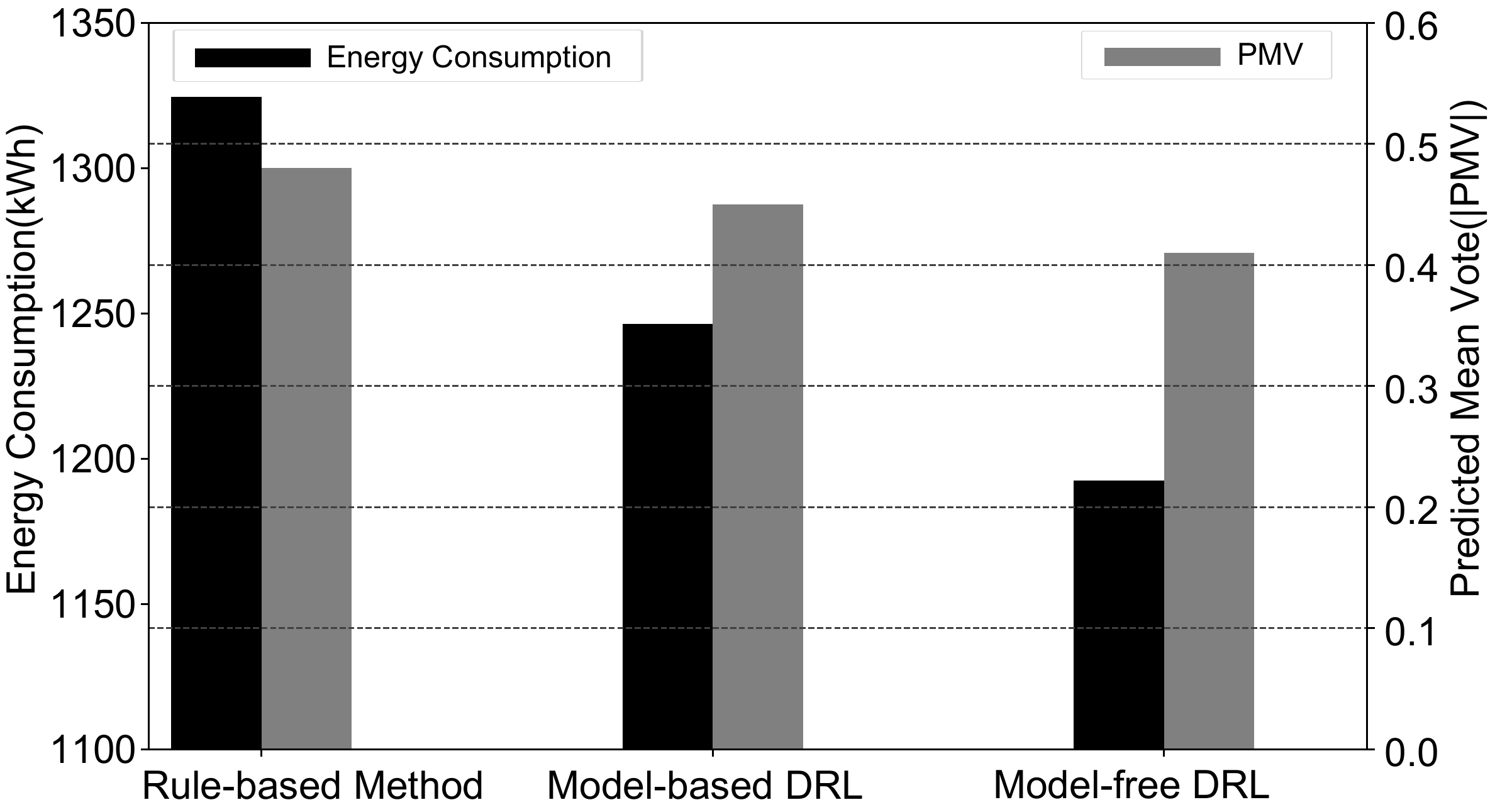}
% 		 \includegraphics[width=2.2in,height=1.35in,angle=0]{samples/ensemble_motivation_bad.pdf}
% 		 \vspace{-0.1in}
		\caption{Random shooting in the model-based DRL method.}
		\label{different_controller}
	\end{minipage}	

\end{figure*}

\section{Related Work}\label{sec:relate_work}
MPC solves an optimal control problem iteratively over a receding time horizon. \cite{beltran2014optimal} proposed an MPC approach for HVAC control, which minimizes energy use while satisfying occupant comfort constraints. A very recent MPC work, OFFICE~\cite{winkler2020office}, proposed a novel MPC framework that optimally manages the trade-off between energy cost and quality of comfort to the building users.
OFFICE uses a gray-box approach, where a parametrized first-principled model is used, and the parameter of the model are dynamically learned and updated over time.  In our case, we use a black-box approach, where the neural network learns from scratch the relationships between inputs and outputs in the system.  Also, the MPC controller used is also different.  While OFFICE uses an interior-point method based on a derivable function to find the optimal solution, we use an MPPI controller, which uses sample noise for the exploration around the default values as a search mechanism to find the best optimization solution.

\textbf{Model-free DRL for HVAC control}. Reinforcement Learning has been applied to many areas \cite{ding2022drlic, wu2019towards, fraternali2020aces, shen2020dmm, shen2019deepapp, liu2020continuous}, e.g., irrigation control~\cite{ding2022drlic}, anomaly detection~\cite{wu2019towards}, sensor configuration~\cite{fraternali2020aces}, Map matching for cellular data\cite{shen2020dmm}, Mobile Application Usage Prediction\cite{shen2019deepapp}. 

In particular, MFRL techniques have demonstrated the potential optimal HVAC controls. In MFRL schemes, the agent learns the policy by extensively trial-and-error interaction with the environment. \cite{park2020hvaclearn} leveraged RL to calculate thermostat set-points to balance between occupant comfort and energy efficiency. 
\cite{zhang2018practical} implemented and deployed a DRL-based control method for
radiant heating systems in a real-life office building. A holistic building control accounting for HVAC, lighting, window opening and blind inclination was studied using branching dueling Q-network (BDQ) in~\cite{ding2019octopus}. However, practical application of RL was limited by its sample complexity, i.e. the long training time required to learn control strategies, especially for tasks associated with a large state-action space. Le et al. \cite{le2021deep} propose a control method of air free-cooled data centers in tropics via DRL. Vazquez-Canteli et al. \cite{vazquez2020marlisa} develop a multi-agent RL implementation for load shaping of grid-interactive connected buildings. Zhang et al. \cite{zhang2018practical, zhang2018deep} implement and deploy a DRL-based control method for radiant heating systems in a real-life office building. Gao et al. \cite{gao2020deepcomfort} propose a deep deterministic policy gradients (DDPGs)-based approach for learning the thermal comfort control policy. Although the above works can improve the performance of HVAC control, they only focused on HVAC subsystem. Gnu-RL~\cite{chen2019gnu} adopted a differentiable MPC policy, which encodes domain knowledge on planning and system dynamics, making it both data-efficient and interpretive. However, they assumed that dynamics of a water-based radiant heating system can be locally linearized. The assumption worked for the problems they considered, but it may not extrapolate to more complex problems like ours.

\textbf{Model-based DRL for HVAC control}.
To reduce sample complexity, researchers have adopted model-based deep reinforcement learning for HVAC control~\cite{zhang2019building}. In this work, they proposed an MBRL approach that learns the system dynamics using a neural network. Then, they adopt MPC using the learned system dynamics to perform control with RS method. MBRL method works well when the action and state dimension is low, like single-zone building. They often cannot achieve the final performance as model-free method when they are applied to high state and action dimensions of multi-zone buildings.

\section{Motivation}
To understand the performance of a state-of-the-art MBRL method \cite{zhang2019building}, we perform a set of simulations in EnergyPlus for a building with five zones. 
All system settings are the same as~\cite{zhang2019building}, except the state and action dimension is higher for the five-zone building, i.e. a multi-zone building instead of a single zone.
A deterministic neural network is adopted to model the building dynamics and a random shooting method is used to choose the best heating and cooling setpoints. 
We also implement a simple MFRL-based method, Proximal Policy Optimization (PPO)~\cite{schulman2017proximal}, for comparison in this preliminary evaluation.
This is also used to gain trust that the simulator is being run correctly, with intuitive results that can be understood. 
Our goal is to study the effect of a multi-zone building to existing model-based and model-free DRL control method. 
Thermal comfort is measured by PMV~\cite{fanger1970thermal}, which should be controlled within the range (-0.7$\sim$0.7). The simulations are conducted with weather data for the month
of January. The building is 463 $m^2$ in Fresno CA. It has windows in all four facades and glass doors in south and north facades. The south-facing glass is shaded by overhangs. For our 5-zone building, the state dimension is 37, including indoor air temperature, humidity, PMV, energy consumption for each zone and related outdoor environmental parameters; and the action dimension is 10, including cooling and heating set points for each zone.

\textbf{Experiment results.} 
Figure~\ref{mf_better_mb} shows the energy-saving performance of model-based and model-free DRL control method with 50$\times10^4$ time-steps of training data.  The reward means the energy-saving performance under the reasonable thermal comfort that is defined in Section~\ref{reward_section}. 
We evaluate the accumulated reward every 2976 time-steps (one month).
The performance of the rule-based method is a straight line, because its reward does not change as the weather data and building environment are deterministic. 

From Figure~\ref{mf_better_mb}, we can see that model-based DRL and PPO method need 7.5$\times$10$^{4}$ and 23.75$\times$10$^{4}$ time-steps to behave better performance than rule-based method. For converge time, the model-based method needs 11.5$\times$10$^{4}$ and the PPO method needs 50$\times$10$^{4}$ time-steps. The model-based method is 4.38$\times$ more data-efficient than PPO method. However, in the long run, the model-free method eventually outperforms the model-based method. It's easy to see that the model-free method is a trial and error method and the performance increases when using more training data. However, in this case, our model-based method cannot achieve the same performance as model-free method as the training data increases.
The model-based method performs well when the action and state dimension is low (e.g., 9 in~\cite{zhang2019building}). 
However, both the building dynamics model and the control method may not be efficient when the state and action dimension is high, like 47 in our 5-zone building.

when we consider such a high state and action space, the current model is intractable to capture the building dynamics accurately and thus the controller will select sub-optimal heating and cooling setpoints. Another reason is that the performance of random shooting method is ineffective as the action space increases.
In order to understand the rationale behind this, we go to the details of existing model-based method and analyze its component from two perspectives - the uncertainty of the building dynamics model and the effectiveness of the control method. 

\textbf{Challenge 1 - Model Uncertainty.}
Neural network models may have epistemic uncertainty, due to the lack of sufficient data to uniquely model the underlying system~\cite{lakshminarayanan2017simple, chua2018deep, cs285, nagabandi2020deep}.
In an MBRL-based HVAC control system, a building dynamics model predicts the next state of the building, given the current state (e.g. current zone temperature) and a control action (e.g. actuators' temperature set-points).
Even a small bias of the building dynamics model may significantly impact the decision of the controller~\cite{chua2018deep, cs285}.
We conduct an experiment to study this uncertainty of the existing building dynamics model. 
We use 8000 historical data points to train the model, and 2000 data points for testing.

Figure~\ref{aleatoric_epistemic_uncertainty} shows the predictive zone temperature as a function of the action performed. 
The x-axis shows the temperature differential between the supply temperature (action) and the zone temperature at time $t$, and the y-axis shows the temperature differential between the zone temperature after and before actuation.
The figure depicts the predicted temperatures of two neural network models and the ground truth.  
These two models have the same architecture and are trained with the same training data, but their training processes start with different initialization states.
In the middle region of Figure~\ref{aleatoric_epistemic_uncertainty}, we have sufficient data, since most of the actions in the historical data do not change the state sharply.
In this region, both models can accurately predict the next state.
However, when the actions intend to change the state much, we do not have sufficient data for training, and the performance of the two models diverges.   
\textbf{Challenge 2 - Controller Effectiveness.} RS generates $N$ independent random action sequences $\left\{ a_{t},...a_{t+H-1}\right\}$, where each sequence $A_{i} = \left \{a_{0}^{i}...a_{H-1}^{i} \right \}$ for $i = 1 ... N$ is of length $H$ action. Given a reward function $r\left ( s,a \right )$ that defines the task, and given future state predictions
$\hat{s}_{t+1}= s_{t}+f_{\theta}\left (\hat{s_{t}},a_{t}  \right )$ from the learned dynamics model $f_{\theta}$, the optimal action sequence $A_{i^{*}}$ is selected as the one with the highest predicted reward: $i^{*} = \operatorname*{arg\,max}_{i}R_{i}    = \operatorname*{arg\,max}_{i}\sum_{t^{'}=t}^{t+H-1}r\left ( \hat{s}_{t^{'}},\hat{a}_{t^{'}} \right )$.
This approach has been shown to achieve success in controlling a single-zone building with learned models. However, when we apply it into a five-zone building, it has numerous drawbacks: it scales poorly with the dimension of both the planning horizon and the action space, and it often is insufficient for achieving high task performance since a sequence of actions sampled at random often does not directly lead to meaningful behavior. 

Figure~\ref{different_controller} studies the energy consumption and thermal comfort of three HVAC control methods, including a rule-based method, a model-based method and a model-free method. 
To eliminate the impact of model uncertainty for the model-based method, we use the ground-truth states of the building as the results of the building dynamics model (i.e. perfect future state prediction).
From the Figure~\ref{different_controller}, we can see that all three methods can meet the requirement of thermal comfort with same level of PMV value (0.48, 0.45, 0.41). The energy consumption of the model-based method is 4.70\% higher than the model-free method. It is caused by RS control, because the building dynamics model used in the model-based method is perfect in this experiment. 

Based on the previous observations, our main goal is to overcome the drawbacks of model uncertainty and controller effectiveness and find a method that is able to match the high performance of model-free methods while having the sample/data-efficiency of model-based methods.

\begin{figure*}[t]
\begin{center}
  \includegraphics[height=2.2in, width=7.0in]{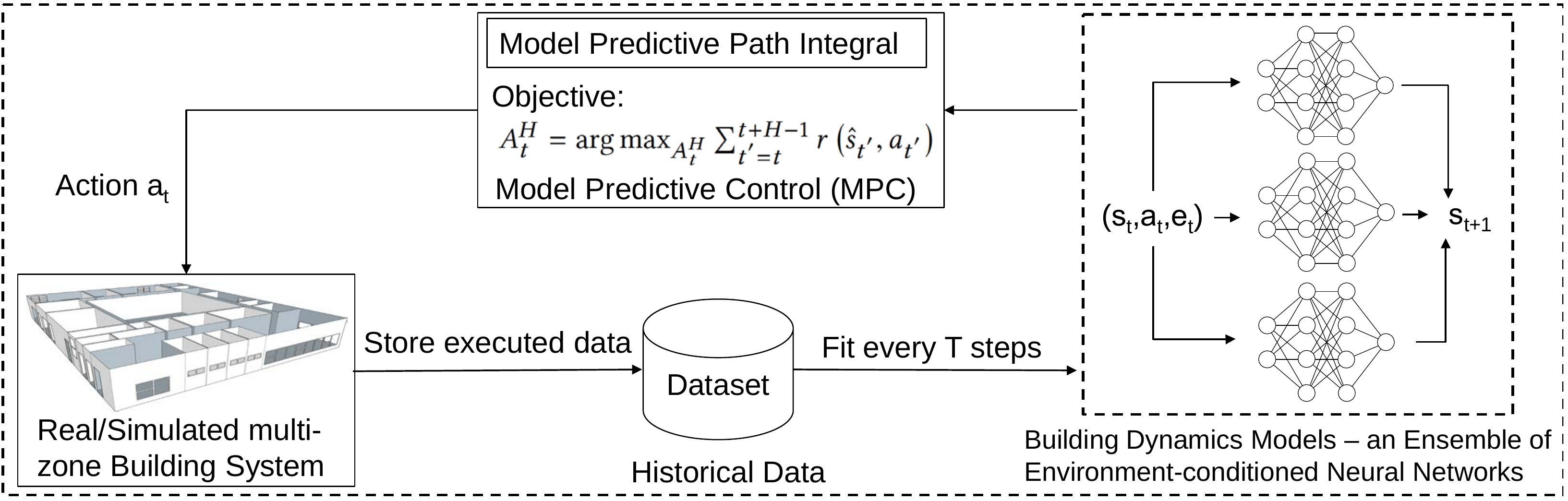}
  \end{center}
  \caption{Overall of the Proposed Building Energy Control Framework}
  \label{framework}
\end{figure*}

\section{Design of \aliasAPP}
In this section, we describe the design of \aliasAPP, including model-based DRL for a multi-zone building control, the building dynamics model and its training details, online control action planning and in-situ update of the building dynamics model.

\subsection{\aliasAPP Overview}
Figure~\ref{framework} shows the overview of \aliasAPP as a model-based DRL control approach \cite{cs285} for multi-zone building HVAC systems. 
At a high level, \aliasAPP includes two key components, i.e., a building dynamics model and a  Model Predictive Path Integral (MPPI) based controller.
Our building dynamics model is built by an Ensemble of multiple Environment-conditioned Neural Networks (ENN).
It takes the current state of the building HVAC system and a specific control action as input, and outputs the next state of the building HVAC system.
Based on the historical data, we train the building dynamics model as a supervised learning process.
With the trained building dynamics model, our MPPI-based controller can evaluate different control actions and find the best control action for next time step, which meets the thermal comfort requirement with minimal energy consumption. 

When we deploy the system in a building, \aliasAPP executes the best control action by setting corresponding actuators every control cycle. 
At the same time, we accumulate building data traces, i.e., the next HVAC state determined by the current HVAC state and the executed control action. 
With the newly collected building traces, we can perform in-situ updating of the building dynamics model periodically (e.g., every week) with a sliding window of 2-months to improve its accuracy, as the seasonality of the data changes during the year. 
One iterative training process takes 25.32 minutes to finish using a laptop with Intel 4-core i7-6700 CPU and Nvidia GTX 960M GPU, and it can be performed in parallel when the current model is being used in the building; thus, the overhead of the iterative training process does not impact the usage of \aliasAPP in real buildings.

\subsection{Model-Based Deep Reinforcement Learning for Multi-zone Building Control}
We extend the current MBRL-based method to multi-zone building HVAC control, including the design of those key components.

\subsubsection{Preliminaries for DRL}
The goal of reinforcement learning is to learn a policy that maximizes the sum of future rewards. At each time step $t$, the controller is in state $s_{t}\in S$, executes some action $a_{t}\in A$, receives reward
$r_{t} = r(s_{t}, a_{t})$, and transitions to the next state s$_{t+1}$ according to some unknown dynamics function $f : S \times A \rightarrow  S$. The goal at each time step is to take the action that maximizes the discounted sum of future rewards, given by $\sum_{t^{'}=t}^{\infty}\gamma^{t^{'}-t}r ( s_{t^{'}},a_{t^{'}} )$, where $\gamma \in  [0,1]$ is a discount factor that prioritizes near-term rewards. Note that performing this policy extraction requires knowing the underlying reward function $r(s_{t}, a_{t})$ that we use for planning actions under the learned model.

In model-based reinforcement learning, a model of the dynamics is used to make predictions, which is used for action selection. Let $f_{\theta}(s_{t},a_{t} )$ denote a learned discrete-time dynamics function, parameterized by $\theta$, that takes the current state $s_{t}$ and action $a_{t}$ and outputs an estimate of the next
state at time $t + \Delta t$. We can then choose actions by solving the following optimization problem:
\begin{equation}
\begin{array}{l}
\left ( a_{t},...a_{t+H-1} \right ) = \operatorname*{arg\,max}_{a_{t},...a_{t+H-1}}\sum_{t^{'}=t}^{t+H-1}\gamma ^{t^{'}-t}r\left ( s_{t^{'}},a_{t^{'}} \right )
\end{array}
\label{optimization}
\end{equation}

In other words, we will pick the action sequence that maximizes the discounted sum of reward of future $H$ time-steps. 
In practice, it is often desirable to solve this optimization at each time step, execute only the first action from the sequence, and then re-plan at the next time step with updated state information.
Such a control scheme is often referred to as model predictive control (MPC), and is known to compensate well for errors in the model.

\subsubsection{State Design}
The state is what the building dynamics model takes as input for the next prediction step. In this study, we separate the state into 2 parts: (a) the building state ($s_{ti}$), which are the state variables that change with our control actions; and (b) the environment state ($e_{ti}$), which are the state variables that do not change with our control actions.

\textbf{Building State ($s_{ti}$)} The building state vector that changes over time $t$ for the $i$th zone consists of the following items: indoor air temperature($^{\circ}$C), indoor air relative humidity (\%), PMV, heating energy consumption (kWh) and cooling energy consumption (kWh).

\textbf{Environment State ($e_{ti}$)} The environment state vector that changes over time $t$ for the $i$th consists of the following items: outdoor air temperature ($^{\circ}$C), outdoor air relative humidity (\%), diffuse solar radiation ($W/m^{2}$), direct solar radiation ($W/m^{2}$), solar incident angle ($^{\circ}$), wind speed (m/s), wind direction and occupancy flag (0 or 1). The occupancy flag is an indicator to detect whether there are people in the $i$th zone, and it is the only element in the vector that changes per zone.

Taking our 5-zone building as an example, the state dimension is 37 including the building, and environment state variables.  The Min-max normalization is used to convert each item to a value within 0-1.

\subsubsection{Action Design}
\label{action_section}
The action vector ($a_{ti}$) shows the actuation variables used by the controller to control the building state ($s_{ti}$). The action state vector that changes over time $t$ for the $i$th zone consists of the following items: cooling temperature set-point and the heating temperature set-point (both in $^{\circ}$C).
Given the current state ($s_{ti}$ and $e_{ti}$) and action ($a_{ti}$), we want the controller to find the most suitable action combinations ($a_{(t+1)i}$) for all the zones to balance energy consumption and thermal comfort metrics.
The action dimension is 10 in our five-zone building.

\subsubsection{Reward Design}
\label{reward_section}

The reward function controls the optimization parameters that want to be maximized when the agent performs an action ($a_{ti}$) to transition from the building state $s_{ti}$ to $s_{(t+1)i}$. 
Both thermal comfort and energy consumption should be incorporated. 
The reward function is defined as follows:
\begin{equation}
  \begin{aligned}
    R = -\sum_{i=1}^{N} \left (  \rho Norm(  \left |  PMV_{i} \right |  )    +   Norm(E_{i})   \right ),
  \end{aligned}
  \label{reward}
\end{equation}

\noindent

where $E$ is heating and cooling energy consumption for each zone, we use Fanger’s formula for the Predictive Mean Vote (PMV)~\cite{fanger1970thermal} to estimate comfortable temperature bounds for the ``standard'' occupant within the current seasonal conditions, as defined by ASHRAE standard 55~\cite{ASHRAE}. The maximum high/low end of the comfort range for Class C environments has PMV values of +/- 0.7. $\rho$ is used to balance the relative importance between energy consumption and thermal comfort. 
We use $\rho$ = 4 during occupied periods and 0.1 during unoccupied periods since the range of human comfort and energy consumption is different during occupied and unoccupied periods. The reward evaluates the actions to meet the requirement of thermal comfort of all the occupants in the building. 
$N$ is the number of zones.
In the following sections, we will remove the $i$ index for each zone to simplify the notation.

\subsection{Learning the Building Dynamics}
We require a parameterization of the building dynamics model that can cope with high-dimensional state and action spaces, and the complex dynamics of a multi-zone building. 
Therefore, we represent the dynamics function $\hat{f_{\theta}}\left (s_{t},a_{t}  \right )$ as a multi-layer neural network, parameterized by $\theta$. 
This function outputs the predicted change in state that occurs as a result of executing action $a_{t}$ from state $s_{t}$, over the time step duration of $\Delta t$. 
Thus, the predicted next state is given by: $\hat{s}_{t+1}= s_{t}+\hat{f_{\theta}}\left (s_{t},a_{t}  \right )$. 
While choosing too small of a $\Delta t$ leads to too small of a state difference to allow meaningful learning, increasing the $\Delta t$ too much can also make the learning process more difficult because it increases the complexity of the underlying continuous-time dynamics. 

\begin{figure}[t]
\begin{center}
     \includegraphics[height=1.1in, width=3.5in]{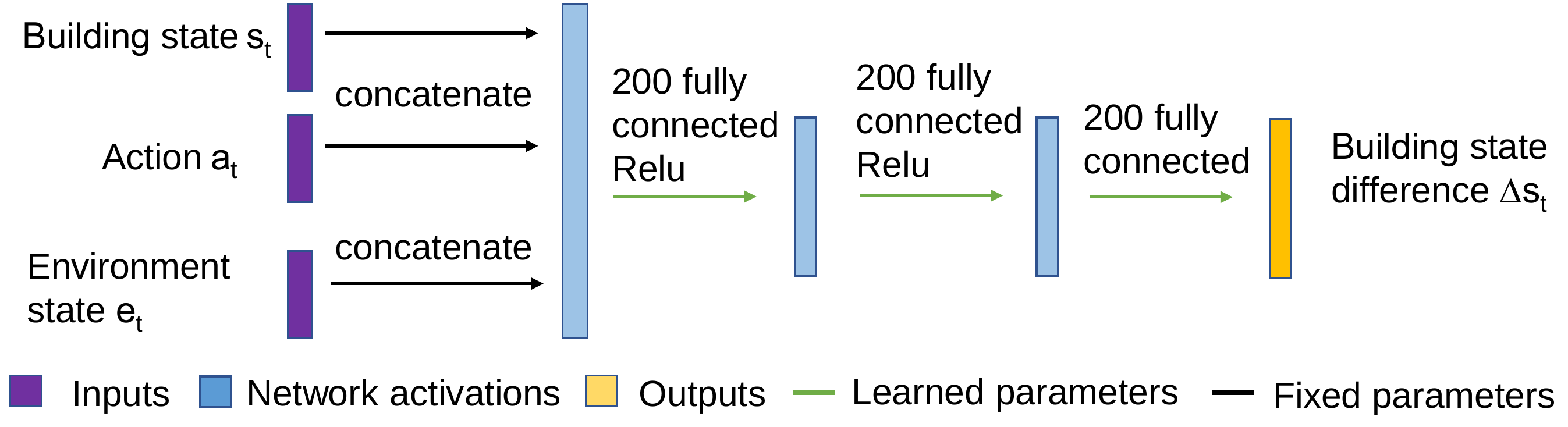}
\end{center}
  \caption{Environment-conditioned neural network for our Building Dynamics Model. }
  \label{model}
\end{figure}

\subsubsection{Environment-conditioned Neural Network Architecture} We define a neural network model $\hat{f_{\theta }} ( s_t,a_t )$ for the building dynamics. In order to make the model achieve both good predictive accuracies and tractable computational optimization, we propose a simple and highly effective method for incorporating environment information. We formulate an environment-conditioned dynamics model $\hat{f_{\theta}} (s_t,a_t,e_t)$ that takes as input not only the current building state $s_t$ and action $a_t$, but also the current environment state $e_t$. The model architecture is shown in Figure~\ref{model}. The building state vector $s_t$, the action vector $a_t$ and the environment state vector $e_t$ are concatenated together and then are passed through two hidden layers and a final output layer. As opposed to a straightforward outputting of all the related states (building and environment), we produce a prediction of building state difference $\Delta \hat{{s}}_{t}$. This reduces the burden of the model to learn the changes in the environment that are not necessary. We provide the ground truth value for environment state, e.g., weather data and occupancy~\cite{chen2019gnu}.

\subsubsection{Weighted Ensemble Learning}
As prior work \cite{chua2018deep, cs285} has shown, capturing epistemic uncertainty in the network weights is important in model-based RL, especially with high capacity models that are liable to over-fit to the training set and extrapolate erroneously outside of it. To solve epistemic uncertainty, we propose a weighted ensemble learning algorithm, which approximates the posterior $p(\theta|D)$ with a set of $M$ models, each with parameters $\theta_{i}$. For deep models, it is sufficient to simply initialize each model $\theta_{i}$ with a different random initialization $\theta_{i}^{0}$ and use different batches of data $D_{i}$ at each training step. 

\begin{figure}[t]
\begin{center}
  \includegraphics[height=1.2in, width=3.7in]{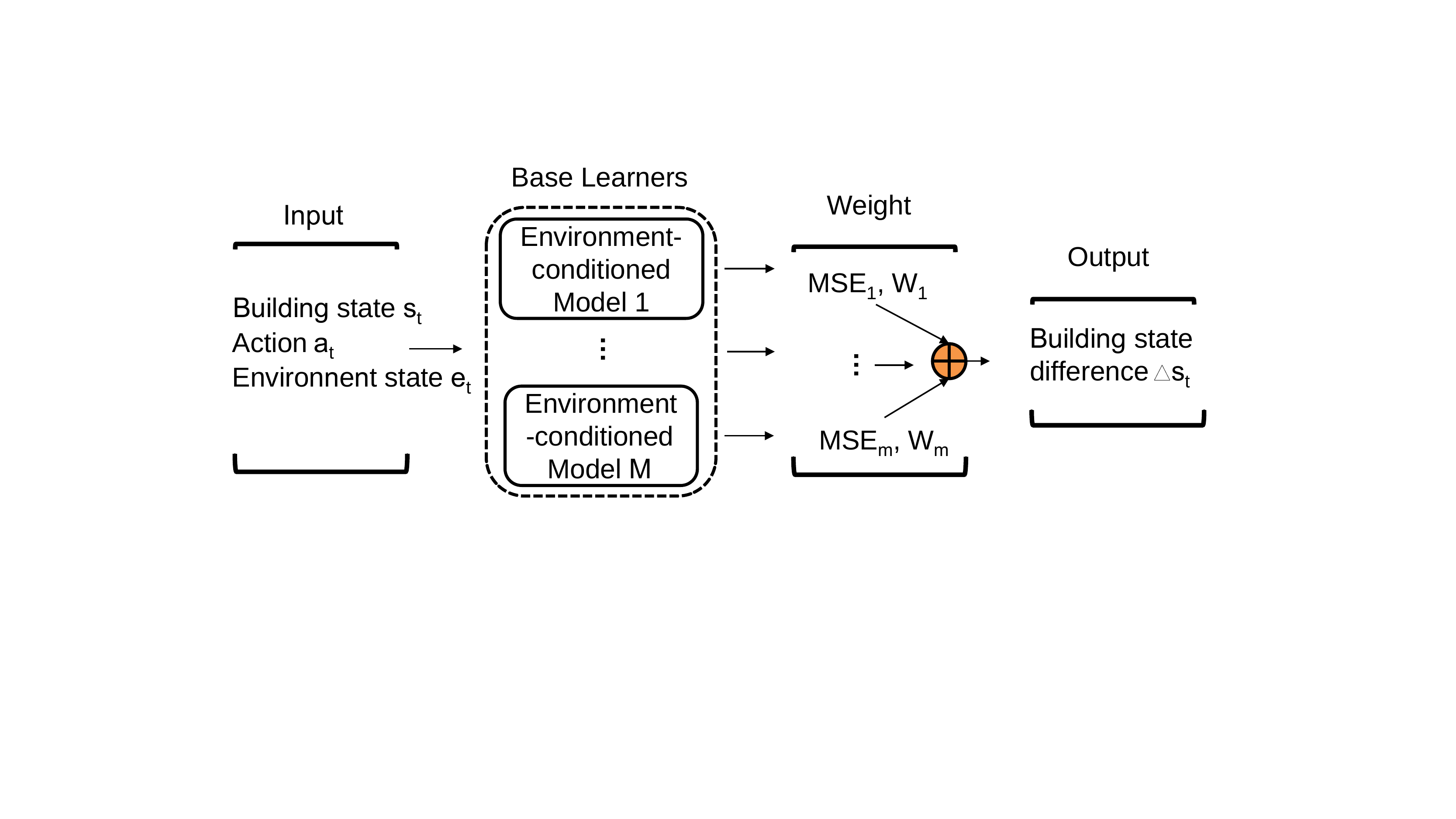}
  \end{center}
  \caption{Weighted Ensemble Learning for our Building Dynamics Model.}
  \label{weighted_ensemble}
\end{figure}

We have $M$ environmental-conditioned models as shown in Figure \ref{weighted_ensemble}. The input for all $M$ models is the same and it includes the building and environment states and actions. To evaluate the performance of each model, we calculate the mean square error ($MSE$) of the past $C$ timesteps (4 in our case) for each model compared to the ground truth for $N$ states using Equation \ref{mse_for_model}. The notation for ensemble algorithm is shown in Table \ref{signal_notation}.
\begin{equation}
\begin{array}{l}

MSE =\sum_{i=1}^{C}\sum_{j=1}^{N}\phi ^{C} \left | f_{\theta }\left ( s_{i,j},a_{i,j} \right ) - \hat{f}_{true}\right |^{2}\\

\end{array}
\label{mse_for_model}
\end{equation}

\begin{table}[t]
  \renewcommand\arraystretch{1.2}
  \small
  \caption{Ensemble Parameters}
  \centering
  \begin{tabular}{|c|c|}
    \hline
      Symbol &   Description\\
    \hline
    N   &number of state \\
    \hline
    $\gamma$  & discount factor  \\
    \hline
     M  & the number of ensemble model \\
    \hline
     E  & number of Ensembler  \\
     \hline
     W  & weight for Ensembler  \\
     \hline
     P  & current state prediction  \\  
    \hline
     C  & past 4 timesteps  \\  
    \hline
    MSE  & model square error  \\  
    \hline
  \end{tabular}
  \label{signal_notation}
\end{table}

We introduce a temporal discount factor $\phi$ (0.9 in our case) that is used to evaluate how important past model error to the current model error. The temporal discount factor is a value between 0 and 1 since recent prediction cases are more important to the performance of current prediction. After we have the $MSE$ for each model of past $C$ timesteps, we first normalize the $MSE$ to 0-1 scale. $Norm(x)$ is a normalization process, i.e., $Norm(x) = (x - x_{min} )/(x_{max} - x_{min})$. Then we calculate the weight ratio $W$ for all models by Equation \ref{weight}.

\begin{equation}
\begin{array}{l}

W = \frac{1-Norm\left ( MSE_{i}   \right )  }{\sum_{i=1}^{M}\left ( 1-Norm\left ( MSE_{i}   \right ) \right )}

\end{array}
\label{weight}
\end{equation}

The sum of all model's weight is 1. After that, we leverage Equation \ref{next_state} to predict the next state. 

\begin{equation}
\begin{array}{l}

s_{t+1} = \sum_{i=1}^{M}W_{i}f_{\theta_{i} }\left(s,a \right)

\end{array}
\label{next_state}
\end{equation}

This allows our method to dynamically adjust the weights in aggregating the $M$ models ($M$=5 in our case) during the prediction. As the states result in unequal prediction accuracy, our method is more robust against this variance.

\subsection{Training the Building Dynamics Model}

In this section, we illustrate how we pre-process training data, and train the proposed ENN model.

\subsubsection{Data Collection} We collect the training dataset ${D}(s_{t}, a_{t}, s_{t+1})$ by executing the rule-based controller at each time step, and recording the resulting data $\tau = (s(0),a(0),s(1),a(1), ... ,s(T-2),a(T-2),s(T-1)$)) of length $T$. We note that these data are very different from the data the controller will end up executing when planning with this learned dynamics model and a given reward function $r(s_{t},a_{t})$ (Section \ref{controller}), showing the ability of model-based methods to learn from off-policy data.

\subsubsection{Data Preprocessing} We slice the collected data $\left \{ \tau  \right \}$ into training data inputs $ \left ( s_{t},a_{t} \right )$ and corresponding
output labels $s_{t+1}-s_{t}$. 
In building HVAC control, states can be temperature, humidity ratio, energy consumption, etc. 
These measurements have various ranges and the weights of the losses will be different if we feed the raw values directly to train the neural network model. 
Thus, we subtract the mean of the states/action and divide by the standard deviation $x' = \frac{x-\bar{x}}{\sigma (x)}$, where x stands for state or action.

\subsubsection{Training the ENN Dynamics Model}\label{one_episode} ENN model consists of an ensemble of models. To make sure the models behave differently on the same dataset $D$, we randomly initialize model parameter $\theta _{1},\theta _{2},..., \theta _{M}$ for all the dynamics models and use different batches of data $D$ at each training step. We train the dynamics model $\hat{f_{\theta}}\left (s_{t},a_{t}  \right )$ using stochastic gradient descent \cite{robbins1951stochastic, deng2022distpro} by minimizing the Mean Square Error (MSE) between predicted delta observation and ground truth delta observation as follows:
\begin{equation}
\begin{array}{l}
\varepsilon \left ( \theta  \right ) = \frac{1}{D} \sum _{(s_{t},a_{t},s_{t+1})\in D} \frac{1}{2} \left \| (s_{t+1}-s_{t} \right) - \hat{f_{\theta}}\left (s_{t},a_{t}  \right )\| ^2

\end{array}
\label{data_train}
\end{equation}

We use 5-year weather data from Fresno, CA and Chicago, IL for the ENN model training and a completely different one-year for testing in this study. 
We provide the ENN model with ground truth information on future environment state, i.e. weather and occupancy~\cite{chen2019gnu}. In our implementation of ENN, we use the Adam optimizer~\cite{kingma2014adam} for gradient-based optimization with a learning rate of $10^{-3}$.
We train the ENN model with a batch size of 512 and a discount factor $\gamma$ = 0.99. The number of epochs is 40. Each dynamics model consists of a neural network of two fully-connected hidden layers of size 200 with relu being nonlinear and a final fully-connected output layer. 
The weights and biases are initialized using the Xavier initialization process~\cite{glorot2010understanding}. The number of samples for MPC controllers (RS, CEM, and MPPI) is 1000. The control cycle (timestep) is 15 minutes that is widely used in classic HVAC control~\cite{avci2013model}. We achieve convergence by 4.75x10$^{4}$ time-steps as explained in Section~\ref{convergence-section}.

\subsection{Online Control Action Planning}\label{controller}
In our method, we use online planning with MPC to select actions via our model predictions. Given the building state $s_{t}$ at time $t$, the prediction horizon $H$ of the MPC controller, and an action sequence $a_{t:t+H}= \left \{ a_{t},...,a_{t+H} \right \}$, the proposed ENN model $\hat{f_{\theta}}\left (s_{t},a_{t}  \right )$ produces a prediction over the resulting data $ s_{t:t+H}$. At each time step $t$, the MPC controller applies the first action $a_{t}$ of the sequence of optimized actions  $A_{t}^{H}= \operatorname*{arg\,max}_{A_{t}^{H}}\sum_{t^{'}=t}^{t+H-1}r\left ( \hat{s}_{t^{'}},a_{t^{'}} \right )$. We adopt the MPPI control method~\cite{williams2017information} to compute the optimal action sequence.

\textbf{Model Predictive Path Integral (MPPI) Controller}.
MPPI control method has been applied to autonomously control a vehicle and get good performance. 
MPPI is an importance-sampling weighted algorithm and considers an update rule that more effectively integrates a larger number of samples into the distribution update. As derived by recent model-predictive path integral work \cite{williams2017information}, this general update rule takes the following form for time step $t$, from each of the $K$ predicted trajectories:

\begin{equation}
\begin{array}{l}

a _{t}^{i+1}  =a_{t}^{i} + \sum_{k=1}^{K}\omega ( \varepsilon ^{k} ) \epsilon _{t}^{k} 

\end{array}
\label{mppi}
\end{equation}

Where $\omega$ is the importance-sampling weight for each trajectory and $\epsilon$ is the noise for exploration. The action for timesteps $t$ of $(i+1) th$ trajectory is the sum of the action for timesteps $t$ of $ith$ trajectory and the noise-weighted average over sampled trajectories.

As shown in the algorithm \ref{MPPI_algo}, an initial control sequence is done either by initializing the input buffer with zeros or by using a secondary controller such as rule-based method and using its inputs as the initial control sequence. We first sample $H$ noise from a normal distribution. Then, we compute $K$ trajectories for $H$ finite horizon with Brownian motion. For each trajectory generated, a cost is computed and stored in memory (line 2-7).

\begin{algorithm}[t]
    \caption{MPPI Controller}
    \label{MPPI_algo}
    \KwIn{
    ENN dynamics model $\hat{f_{\theta}} (s_{t},a_{t} )$\;
    K: Number of samples, H: Length of horizon\;
        $ ( a_{0},a_{1},...a_{H-1} )$: Initial control sequence\;
        $\lambda$:Control hyper-parameter \;
    }
    \KwOut{The control sequence $a_{t:t+H}$ \;}
    $s_{0} \leftarrow  GetStateEstimate()$ \;

    \For{k = 0,1,...,K -1 }
    {
        s $\leftarrow s_{0}$\;
        Sample noise  $ \varepsilon^{k} =  \{  \epsilon_{0}^{k},\epsilon_{0}^{k},...\epsilon_{H-1}^{k}  \}  \sim \mathbb{N} (\mu ,\sigma  )$ \;
        
        \For{t = 1,...,H }
        {
          $s_{t} \leftarrow  \hat{f_{\theta}} (s_{t-1},a_{t-1}+\epsilon _{t-1}^{k}   )$  \;
          $ Cost ( \varepsilon ^k  ) \mathrel{+}= -reward $ defined by equation \ref{reward} \; 
        }
    }
    $\beta \leftarrow min_{k}[   Cost ( \varepsilon ^k  )  ]$ \; 
    $\eta \leftarrow \sum_{k=0}^{K-1}exp ( -\frac{1}{\lambda } ( Cost ( \varepsilon ^k ) - \beta ) )$ \; 
    \For{k = 0,1,...,K -1  }
    {
        $\omega ( \varepsilon ^{k} )\leftarrow \frac{1}{\eta }exp( Cost ( \varepsilon ^k ) - \beta )$\;
    }    
    \For{t = 0,1,...,H -1  }
    {
        $a _{t}^{*}  =a _{t} + \sum_{k=1}^{K}\omega ( \varepsilon ^{k} ) \epsilon _{t}^{k} $;
    }       
    SendToActuators($a_{0}$)\;
    \For{t = 0,1,...,H -1 }
    {
        $a_{t-1}  = a_{t}$;
    }       
    $a_{t-1} = Initialize(a_{t-1})$;

\end{algorithm}

In model predictive control, optimization and execution take place simultaneously: a control sequence is computed, and then the first element of the sequence is executed. This process is repeated using the un-executed portion of the previous control sequence as the importance sampling trajectory for the next iteration. In order to ensure that at least one trajectory has non-zero mass (i.e., at least one trajectory has a lowest cost), we subtract the minimum cost of all the sampled trajectories from the cost function (line 9). Note that subtracting by a constant has no effect on the location of the minimum. In the second loop, we get the noise weighted average over $K$ sampled trajectories (lines 10-11). The third loop computes an optimal input sequence using least cost of the trajectories for $H$ finite horizons (lines 12-13). The top of the stack value is given to the actuators (line 14). After that, the whole input control sequence is left shifted by 1 (lines 15-16). To maintain the length of buffer, $a_{init}$ is appended to the input control sequence (line 17). The states are then updated from the ENN model.

\subsection{Putting It All Together}
We summarize the working flow of \aliasAPP as follows. We first gather historical dataset $D$ using a rule-based policy and randomly initialize model parameter $\theta _{1},\theta _{2},..., \theta _{M}$ for ENN. Then we train the ENN model using this dataset by Equation \ref{data_train}.  
Finally, we deploy the learned ENN model and our MPPI controller in the real building for HVAC control. 

For one control execution, we first obtain the current building state from sensors (e.g., zone temperature from a temperature sensor). 
After that, the best action sequence is sampled by MPPI controller with $H$ horizon and the state is propagated by ENN model by solving the optimization problem defined in Equation~\ref{optimization}. 
We execute the first action of the optimal action sequence in the building by setting corresponding actuators. 

When \aliasAPP is running in the building, we can also collect building operation data, which is composed of control action execution records $D(s_{t},a_{t},s_{t+1})$, including current state, control action, and next state. 
We add the newly collected data into a sliding window for two months of data and train the ENN model from scratch again. 
We use a sliding window to adapt to the seasonality of the data, especially weather data.  
We randomly divide the training data set into a set of batch and update the weight through forward and backward propagation by feeding the data into the model. 
This process is called one epoch training after traversing all the batch of data. 
We will repeat this process for multiple epochs (40 in our current implementation) until the model converges.  
This is an iterative in-situ updating process to improve the accuracy of our building dynamic model.

\section{Evaluation}\label{sec:evaluation}
In this section, we conduct a variety of experiments in EnergyPlus to evaluate the performance of \aliasAPP and three baselines by a set of performance metrics.

\begin{figure}[t]
\begin{center}
  \includegraphics[height=1.4in, width=3.4in]{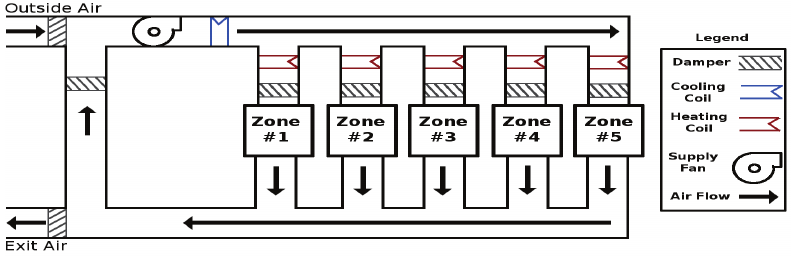}    
\end{center}
  \caption{HVAC Single Duct VAV Terminal Reheat Layout \cite{beltran2014optimal}.}
  \label{hvac_system}
\end{figure}

\subsection{Platform Setup}
\textbf{Building Example and its Dynamics Model in EnergyPlus}
In this work, we evaluate the performance of \aliasAPP in a building of 463 $m^2$ at Fresno, California. It is a single floor building of 5 thermal zones. There are windows on all 4 facades. The HVAC system we modeled is a single duct central cooling HVAC with terminal reheat as shown in Figure \ref{hvac_system}. The process begins at the supply fan in
the air handler unit (AHU), which supplies air for the zone. The supply
fan’s air first goes through a cooling coil, which cools the air to the minimum required temperature needed for the zone. Before air enters a zone, the air passes through a variable air volume (VAV) unit that regulates the amount of air that flows into a zone. Terminal reheat occurs when the heating coil increases the temperature before discharging air into a zone. A discharge setpoint temperature is selected for each zone and the VAV ensures that the air is heated to this temperature for each zone. The air supplied to the zone is mixed with the
current zone air, and some of the air is exhausted out of the zone to maintain a constant static pressure. The return air from each zone is mixed in the return duct, and then portions of it may enter the economizer.

Since we cannot conduct control experiments in the real building, we leverage a building model in EnergyPlus version 8.6 and conduct simulations with Typical Meteorological Year 3 (TMY3) weather data. In our implementation, the AHU set-point is set by default EnergyPlus control logic, and we only control the heating and cooling set-point in the VAV box.

EnergyPlus has been widely used to evaluate the HVAC control algorithm~\cite{zhang2018practical, kumar2021building, chen2019gnu, zhang2019building, ding2019octopus}. There are four reasons why we choose EnergyPlus. First, we do not have one real building that allows us to conduct experiments. \aliasAPP could be deployed in a real building after we finish the ENN model training. Second, it is convenient to generate enough historical training data of rule-based method to train the ENN model. Third, in order to compare with a model-free DRL, we need a significant training data set to train these models since MFRL is not sample efficient.  In our case, we need 5200 days (14+ years) of training data, which is unreasonable to obtain from real buildings. Finally, it is easy for us to evaluate the performance of different control algorithms under different locations, seasons and weather profiles.

\textbf{\aliasAPP System Components}
As shown in Figure \ref{framework}, \aliasAPP system includes two main parts: the building dynamics model ENN and the MPPI controller. We also need to store the newly collected building operation data for in-situ update of the building dynamics model.
All these three components are all implemented in Tensorflow, which is an open-source machine learning library in Python. 
We use the building control virtual testbed (BCVTB)~\cite{wetter2011co} for establishing a connection between EnergyPlus and \aliasAPP. We execute the control action by setting the temperature to a specific set point for each zone of our EnergyPlus building model during each control cycle. 

\begin{table}[t]
  \renewcommand\arraystretch{1.2}
  \small
  \caption{Parameter Settings in \aliasAPP}
  \centering
  \begin{tabular}{|c|c|}
    \hline
      Batch Size &   512\\
    \hline
    Time Step for Control &  15min\\
    \hline    
    Train/Validation Split Ratio   &80\%/20\% \\
    \hline
    Discount Factor $\gamma$  & 0.99  \\
    \hline
    Learning Rate  & 0.001  \\
    \hline    
    Number of Hidden Layers  & 2  \\
    \hline
    Number of Neurons for Each Layer  & 200  \\    
    \hline
    Number of Data Samples  & 1000  \\
    \hline
    Length of Horizon  & 20  \\
    \hline
  \end{tabular}
  \label{mb2c_parameter}
\end{table}

\subsection{Experiment Setting}
We show the parameter setting of \aliasAPP in Table \ref{mb2c_parameter}. The timestep for HVAC control is 15 minutes. The reason is that 15-minute control cycle is widely used in classic building control \cite{ma2014stochastic, avci2013model}. There may have a benefit in doing finer-grained control. As stated in the manual of EnergyPlus, the building dynamics models may be more accurate for shorter timesteps (10 minutes or less). However, based on the experience of classic control, the length of a control period should be no less than 10–15 min, because switching more frequently than once every 10–15 min can physically damage the normal HVAC equipment, like a heat pump \cite{avci2013model}.

We train the ENN model based on the weather data from two different cities, Fresno, CA and Chicago, IL, due to their distinct weather characteristics. Fresno has intensive solar radiation and large variance in temperature, while Chicago is classified as hot-summer humid continental with four distinct seasons.  

We compare \aliasAPP with the three baselines. We execute these four control methods to control the building HVAC system using the same weather data for simulation.

\textbf{Rule-based Method:} We implement a rule-based method according to our current campus building control policy for training data generation and comparison
evaluation. We assign different zone temperature set-points. Each zone has a separate heating and cooling set-point. The heating set-point is set to 70 $^{\circ}$F, and the cooling set-point to 74 $^{\circ}$F during the warm-up stage. The cooling set-point is limited between 72$^{\circ}$F and 80$^{\circ}$F, and the heating set-point is limited between 65$^{\circ}$F and 72$^{\circ}$F. 

\textbf{Model-free DRL:} We implement Proximal Policy Optimization (PPO) \cite{schulman2017proximal} that is the default reinforcement learning algorithm at OpenAI because of its ease of use and good performance.

\textbf{Model-based DRL with RS:} For the conventional model-based method, we implement the deterministic neural network to model the building dynamics and RS method to choose the heating and cooling setpoints \cite{zhang2019building}. 

\begin{figure}[t]
  \includegraphics[height=2.1in, width=3.3in]{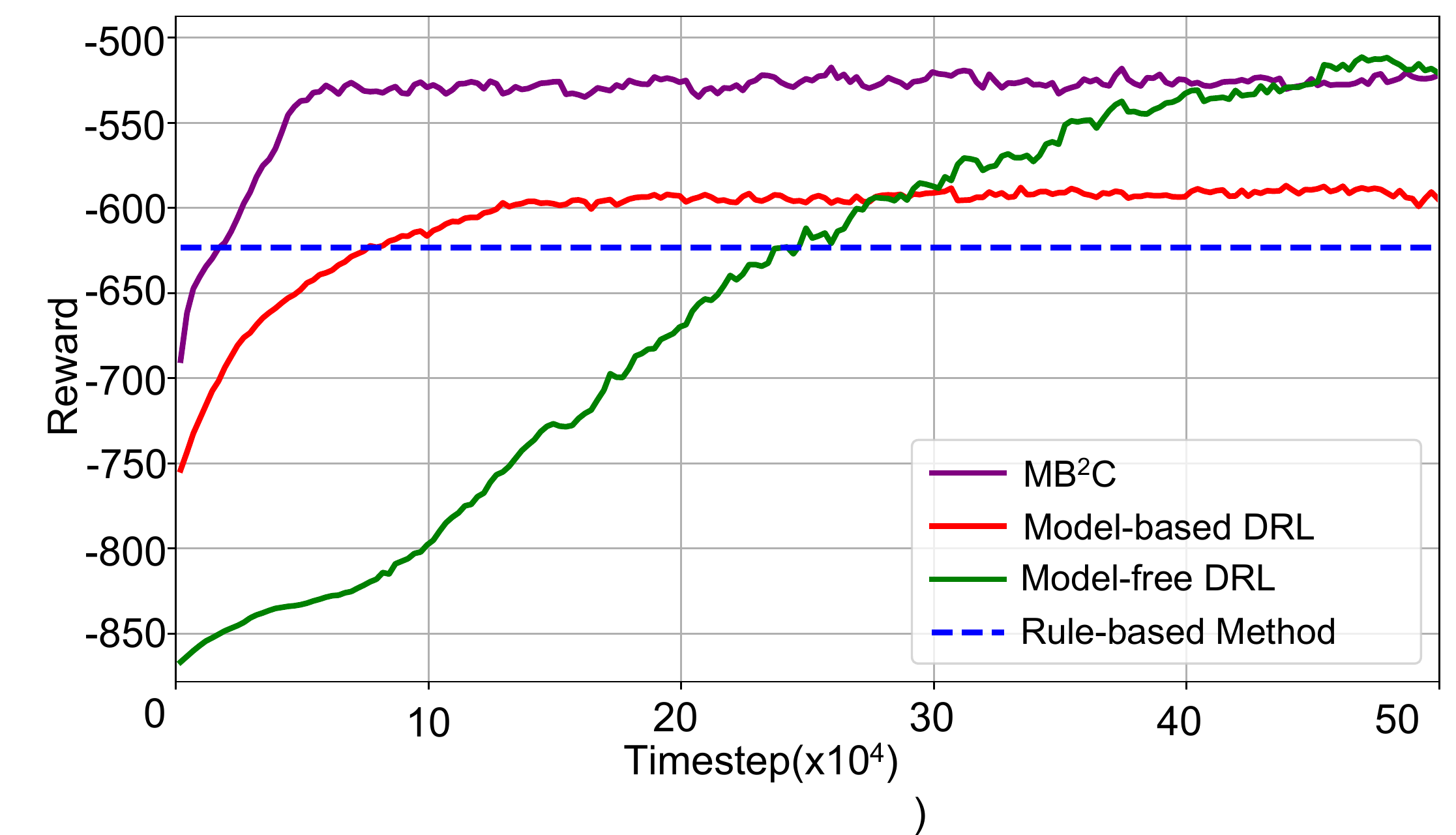}
  \caption{\aliasAPP Achieves both Data-Efficiency and High Performance.}
  \label{reward_convergence}
\end{figure}

\subsection{Experiment Results}
We compare \aliasAPP with the above baselines by a set of performance metrics, including convergence analysis, energy efficiency and thermal comfort. We also study the performance of \aliasAPP, including its daily energy consumption for each zone, the performance gain of its key components, and its parameter setting.

\subsubsection{Convergence Analysis} 
\label{convergence-section}
We first study the data efficiency of \aliasAPP and the other three baselines. For this study, we do not limit ourselves to a sliding window of two months for \aliasAPP, since the MFRL method requires copious amount of training data.
Figure~\ref{reward_convergence} shows that the accumulated reward of four control methods in each episode during a training process.  One episode contains the data collected in one month, corresponding to 2976 time-step. We calculate the reward function every timestep. The reward in Figure~\ref{reward_convergence} is the  accumulated reward of one episode, i.e., the sum of the rewards of 2976 time-steps. 
From the results in Figure \ref{reward_convergence}, we see that the episode reward increases and tends to be stable as the number of training episodes increases. When the episode reward does not change much, it means that we cannot do further to improve the learned control policy and thus the training process converges. 

As indicated in Figure \ref{reward_convergence}, \aliasAPP behaves better than rule-based method after the 1.75$\times$10$^{4}$ time-steps. In this stage, the ENN model is first learned from off-line historical data. Then it can be deployed into real buildings and leverages the MPPI controller for exploration to further improve its performance. The model-based DRL and model-free DRL need 7.5$\times$10$^{4}$ and 23.75$\times$10$^{4}$ time-steps to behave better than rule-based method. \aliasAPP achieves 4.28$\times$ and 13.57$\times$ more data-efficient than model-based DRL and model-free DRL. 

For convergence time, \aliasAPP converges faster than both model-based DRL and model-free DRL.  \aliasAPP needs 4.75$\times$10$^{4}$ and model-based DRL needs 11.5$\times$10$^{4}$ time-steps. The model-free DRL needs 50x10$^{4}$ timesteps. \aliasAPP is 2.4$\times$ and 10.52$\times$ data-efficient than model-based DRL and model-free DRL with the same performance as model-free DRL.

\begin{figure}[t]
  \includegraphics[height=2.1in, width=3.3in]{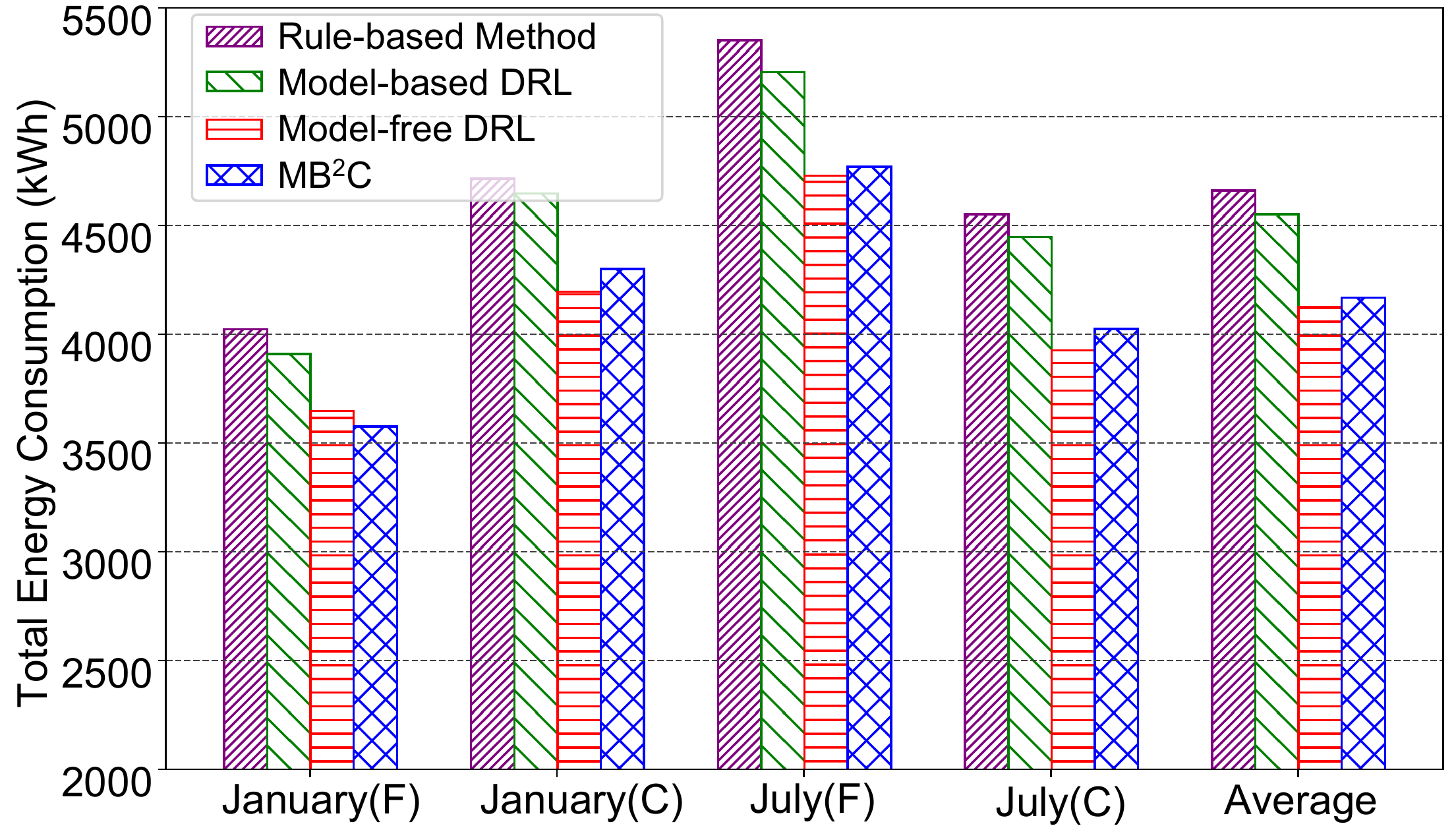}
  \caption{Energy Consumption of \aliasAPP and the Other Baselines.}
  \label{four_method_energy_consumption}
\end{figure}

\subsubsection{Energy Efficiency}
Figure~\ref{four_method_energy_consumption} depicts the energy consumption results of four control methods.
The results reveal that \aliasAPP saves 10.65\% and 8.23\% energy on average, compared with the rule-based method and model-based DRL. Compared with model-free DRL, \aliasAPP achieves comparable performance. 
% In both cities, \aliasAPP achieves similar performance gain. 
\aliasAPP reduces the energy consumption of HVAC by modeling the complex building dynamics accurately and finding better heating and cooling setpoints.

We can also find that for different seasons and cities, the energy consumption is different. In Fresno, the building consumes 4770.04 kWh in July which is 33.39\% more energy than that in January which consumes 3576.07 kWh. The reason is that in July, the outdoor air temperature range at Fresno is 15$^{\circ}$C $\sim$ 42$^{\circ}$C. We have to keep cooling in daylight. However, in January, the outdoor air temperature range at Fresno -1$^{\circ}$C $\sim$ 18$^{\circ}$C. This means that we can use outside air that is already in best range of thermal comfort to save energy.

In Chicago, the building consumes 4300.47 kWh in January that is 6.86\% more energy than in July, because the weather is cold and the outdoor air temperature range in Chicago is -20$^{\circ}$C $\sim$ 15$^{\circ}$C. In July, the outdoor air temperature range at Merced and Chicago is similar, 15$^{\circ}$C $\sim$ 42$^{\circ}$C and 15$^{\circ}$C $\sim$ 40$^{\circ}$C respectively. But the energy consumption in Fresno is 18.53\% higher than the energy in Chicago. The reason is that the average day and night temperature difference for each day is larger than Chicago.

\subsubsection{Thermal Comfort}
Table~\ref{human_comfort} presents the average PMV value for all five zones in January and July under Fresno and Chicago weather data.  All four control methods can maintain the PMV value in the desired range (-0.7$\sim$0.7) for most of the time. The average violation rate of model-based method is 1.97\%, which is a little higher than the other three methods, because the controller tries random actions and some of the actions may lead to bad thermal comfort. 
\aliasAPP achieves a low average violation rate by leveraging more accurate ENN model  and more effective MPPI controller.

\begin{figure}[t]
  \includegraphics[height=2.1in, width=3.3in]{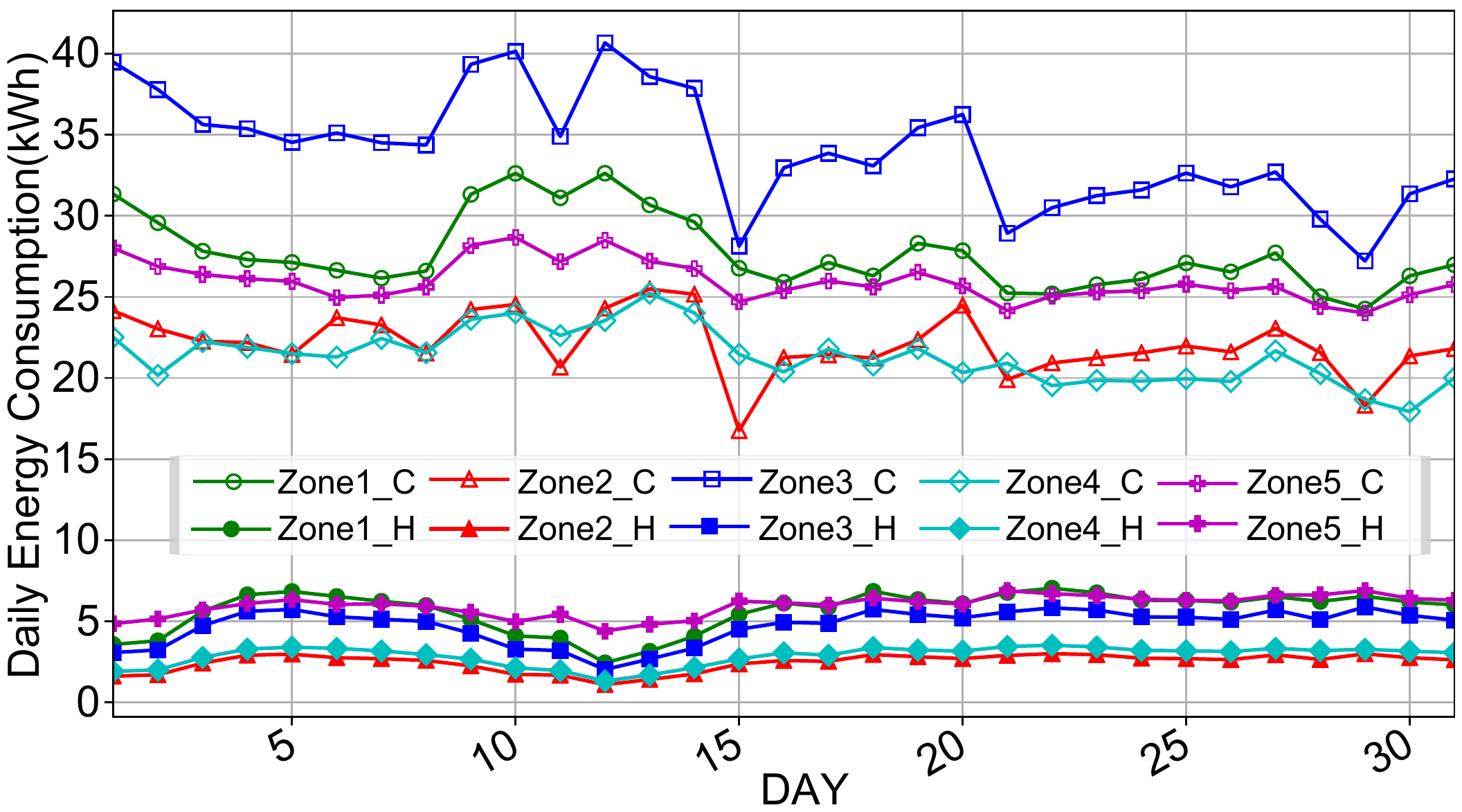}
  \caption{Daily Energy Consumption for Five Zones.}
  \label{fivezone_energy}
\end{figure}

\begin{table*}[t]
  \renewcommand\arraystretch{1.2}
  \small
  \caption{Thermal Comfort Statistical Results for Rule-based, Model-based, Model-free and \aliasAPP Schemes}
%   \vspace{-0.1in}
  \centering
  \begin{tabular}{|c|c|c|c|c|c|c|c|c|c|c|c|}
    \cline{1-11}\multirow{2}{*}{Location}
    &\multirow{2}{*}{Comfort}  & \multirow{2}{*}{Metric} &\multicolumn{2}{c|}{Rule-based\ method}&  
    \multicolumn{2}{c|}{\shortstack{Model-based\ method} }
 
    & \multicolumn{2}{c|}{\shortstack{Model-free\ based method}}&\multicolumn{2}{c|}{\shortstack{\aliasAPP}} \\
    \cline{4-11}
    &  & & January&July&January&July&January&July&January &July\\
    \cline{1-11}
    \multirow{3}{*}{\shortstack{Fresno}}
    & \multirow{3}{*}{\shortstack{PMV}} & Mean & -0.36&-0.20&-0.32&-0.19&-0.11&-0.03&-0.04 &0.13\\
    \cline{3-11}
    &  & Std& 0.26& 0.36  &  0.31& 0.34& 0.15&0.18&0.11 &0.14 \\
    \cline{3-11}				
    &  & Violation rate & 1.22\%& 1.51\%  &  2.12\%& 1.71\%& 	0&0.14\%&0.40\% &0.58\% \\

    \cline{1-11}

    \multirow{3}{*}{Chicago}			
    & \multirow{3}{*}{\shortstack{PMV}} & Mean &-0.17 &-0.30 & -0.26& -0.18& -0.25&0.07 &-0.23 &0.05\\
    \cline{3-11}				
    &  & Std& 0.23&  0.33  &  0.24  & 0.31& 0.17 &0.19  &0.07 &0.20\\
    \cline{3-11}					
    &  & Violation rate &1.20\% &2.04\% & 1.9\% & 2.13\%&0.95\% & 0 &0.46\% &1.23\%\\

    \cline{1-11}

  \end{tabular}
  \label{human_comfort}
%   \vspace{-1em}
\end{table*}

\begin{table}[t]
  \renewcommand\arraystretch{1.2}
  \small
  \caption{Effect of Different Network Architecture}
%   \vspace{-0.1in}
  \centering
  \begin{tabular}{|c|c|c|c|}
    \hline
    \shortstack{Number of\\hidden layers} &\shortstack{Energy Consumption\\(kWh)}
     &  Thermal Comfort  \\
    \hline
    1& 4911 & 0.11\\
    \hline
    2& 4820& 0.15 \\
    \hline
    3 & 4988 & 0.09  \\
    \hline
    4 & 5032 & 0.08  \\
    \hline

  \end{tabular}
  \label{network_architecture}

\end{table}

\subsubsection{Neural Network Architecture}
We explore the effect of different neural network architectures to the energy consumption and thermal comfort using weather data of July in Fresno.
We test four neural networks with different numbers of hidden layers, i.e., 1, 2, 3, and 4. Table \ref{network_architecture} presents the energy consumption and thermal comfort that \aliasAPP can achieve with these four neural networks. From the experiment results, we can see that the neural network with more hidden layers provides better thermal comfort (the result close to 0), but with higher energy consumption. The reason is that the \aliasAPP tries to balance the energy consumption and thermal comfort, the more energy the HVAC system consumes, the more comfort people feel. In \aliasAPP, we set the number of hidden layers to 2, since it can minimize the energy consumption compared to the other three options while meeting the requirement of thermal comfort.

\subsubsection{Daily Energy Consumption for Five Zones}
We analyze the daily energy consumption of \aliasAPP for five zones in July at Fresno. 
As shown in Figure~\ref{fivezone_energy}, we record the heating energy and cooling energy for each zone per day. The top five hollow line symbols record the trend of cooling energy for five zones respectively. The bottom five solid line shows the trend of heating energy for five zones respectively. 
The energy spent by the third zone is higher than the other zones, because the third zone is south-oriented and the sunlight hits into that zone most of the time.

We also see that both heating and cooling occurs in some days, because the day and night temperature difference is large. In the daylight, the average outdoor temperature is 38$^{\circ}$C, and thus we need more energy for cooling. However, at night, the average outdoor temperature is 15$^{\circ}$C, and thus we need some heating air to meet the minimum requirement of thermal comfort (in our simulations we assume an office-like environment with students working at night sometimes). 

\begin{figure}[t]
  \includegraphics[height=2.1in, width=3.3in]{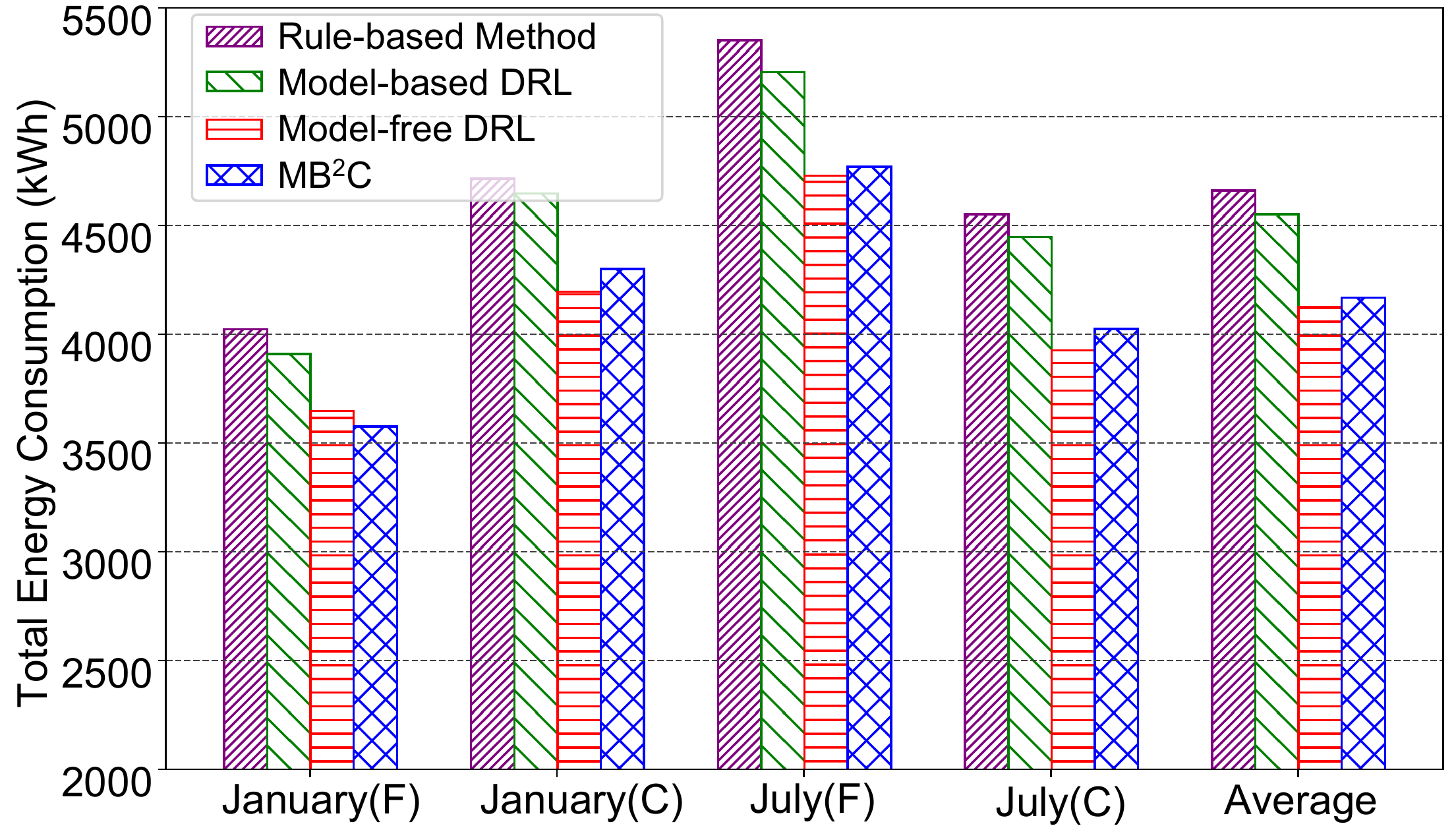}
  \caption{Energy Decomposition.}
  \label{energy_decomposition}
\end{figure}

\subsubsection{Performance Decomposition}
We implement three versions of \aliasAPP with different control methods, i.e., RS (MB\_ENN\_RS), CEM (MB\_ENN\_CEM) and MPPI control method (MB\_ENN\_MPPI). We also compare with the rule-based method and the existing model-based DRL method (MB\_DNN\_RS).

For MB\_ENN\_CEM, we implement (Cross-entropy method) CEM \cite{botev2013cross} controller that begins as the RS method and does this sampling for multiple iterations $m\in \left \{ 0...M \right \}$ at each time step. 
The top $J$ highest-scoring action sequences from each iteration are used to update and refine the mean and variance of the sampling distribution for the next iteration. After $M$ iterations, the optimal heating and cooling actions are selected to be the resulting mean of the action distribution.

Figure~\ref{energy_decomposition} demonstrates the energy consumption of these four methods in two different months and at two different places (Fresno and Chicago). Compared with the rule-based method, MB\_DNN\_RS can only save 2.42\% energy. When the building dynamics model in MB\_DNN\_RS changed to proposed model (MB\_ENN\_RS), 3.34\% more energy can be saved, which illustrates the efficiency of proposed model. When we change the RS method to CEM method and MPPI method with the proposed model, 2.39\% and 4.89 \% more energy can be saved that illustrates efficiency of the MPPI controller.

\begin{figure}[t]
  \includegraphics[height=2.1in, width=3.3in]{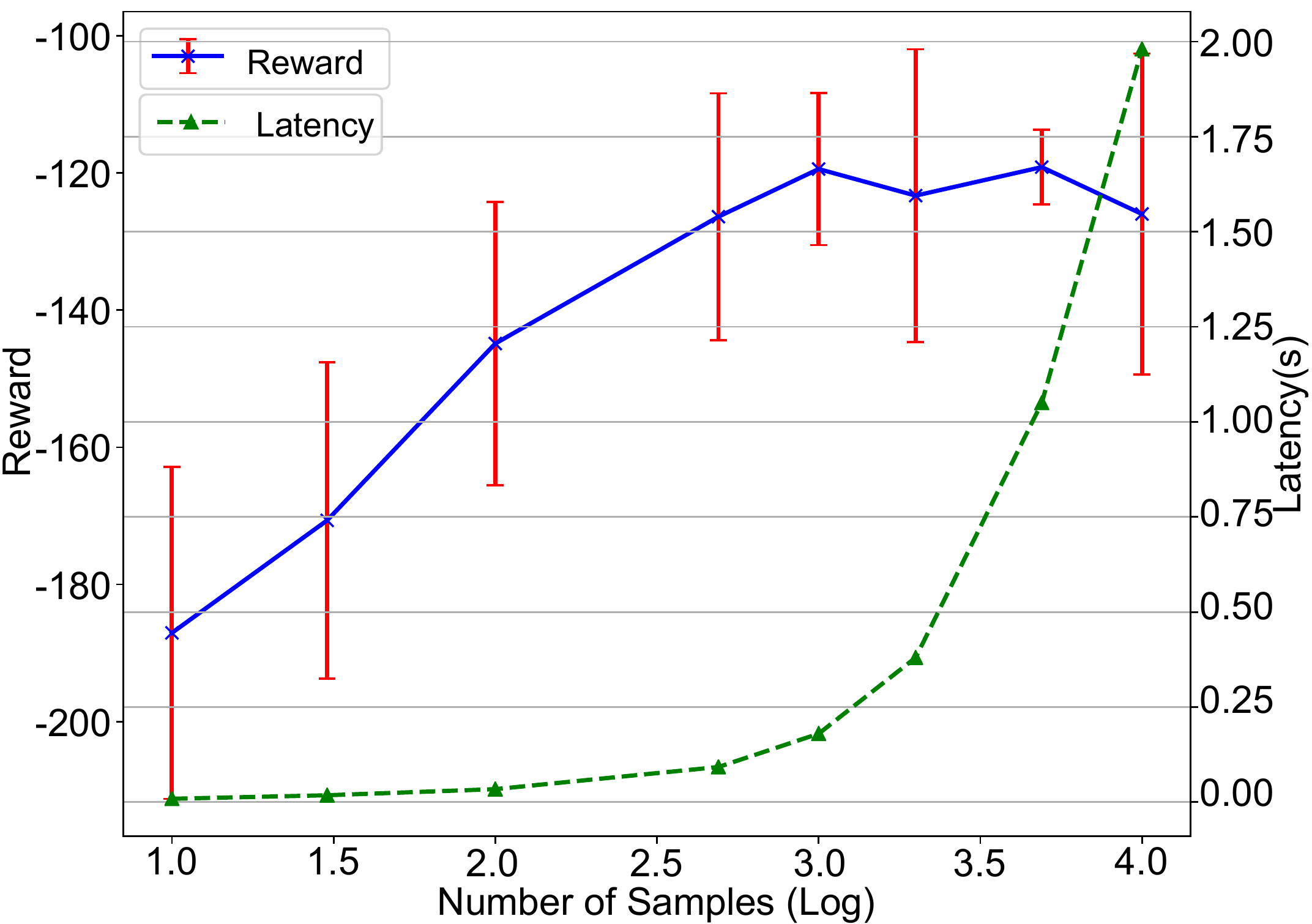}
  \caption{Samples of MPPI Controller.}
  \label{k_value}
\end{figure}

\subsubsection{Parameter Setting}
\aliasAPP has two important parameters that may influence its performance.

\textbf{The Number of Samples in the MPPI Algorithm.}
Figure \ref{k_value} illustrates the performance of the MPPI controller as the number of sample trajectories is changed. We run MPPI controller with ground truth model to investigate the effect of different number of trajectories (10, 30, 100, 500, 1000, 2000, 5000, 10000). We ran 10 times to calculate the mean and standard reward for each number of trajectories. From Figure~\ref{k_value}, we can see that the reward increases quickly as we increase the number of trajectories before 1000 trajectories (power of 3 in the figure). Then it increases slowly after 1000 trajectories, indicating that it is enough for the MPPI algorithm to converge. We also calculate the latency for making one action selection under different number of trajectories. we can see that the latency increases exponentially when trajectories increase. Thus we choose 1000 as the number of trajectories by considering the best reward and lower latency trade-off.

\begin{figure}[t]
  \includegraphics[height=2.1in, width=3.3in]{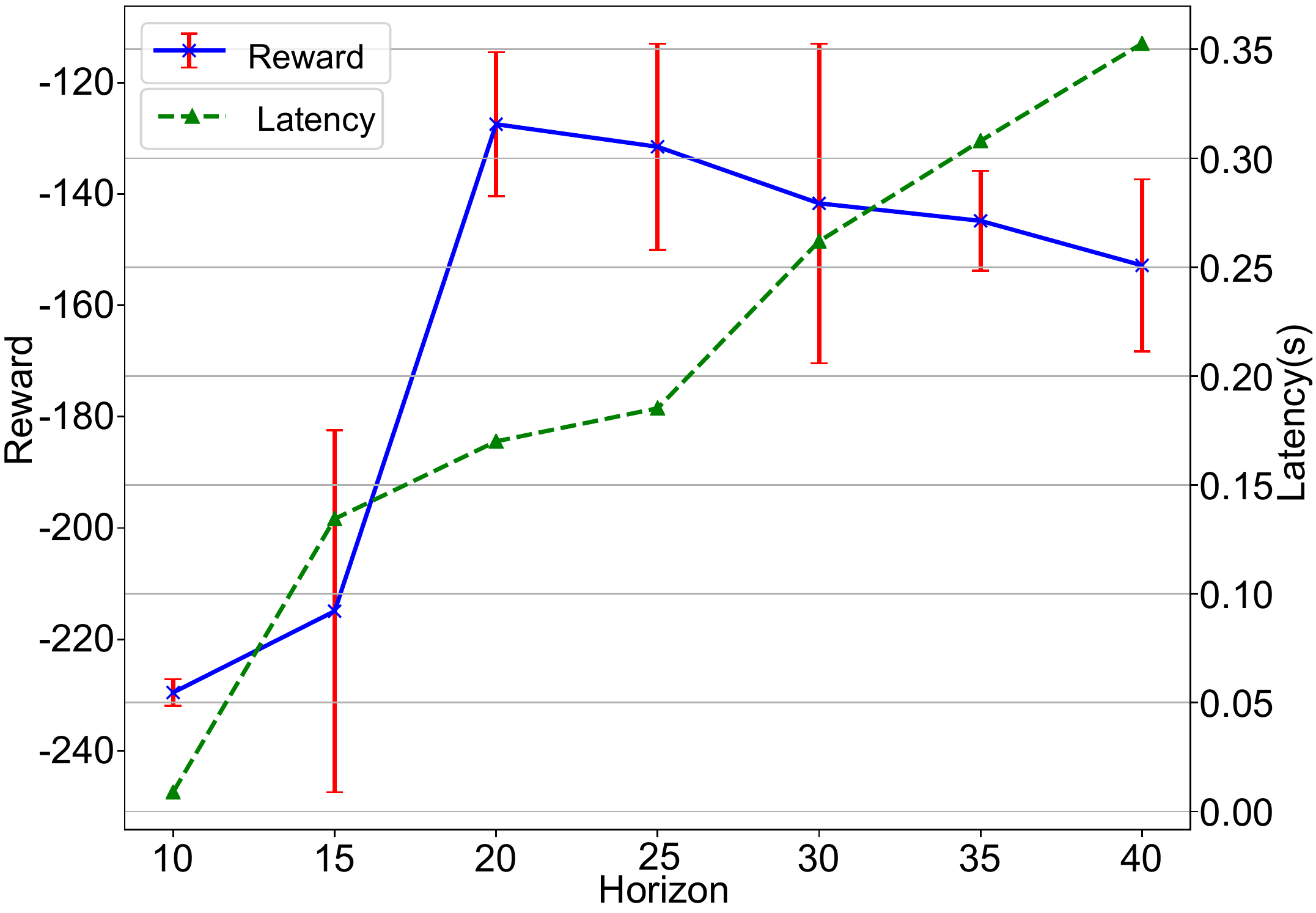}
  \caption{Horizon of MPPI Controller.}
  \label{horizon}
\end{figure}

\textbf{The Length of Horizon in the MPC Process.}
The horizon refers to the number of steps to look ahead in the MPC process. We investigate the effect of different length of Horizon $H$ in Algorithm \ref{MPPI_algo} to the performance of MPPI Controller. From Figure \ref{horizon}, we can see that the reward increases as the length of $H$ increases and achieves the highest reward when $H$ is 20. Then the reward decreases when we continue increasing the length of $H$. The reason is that small horizon results in more greedy actions that may not consider future dynamics. Large horizon produces worse actions since the prediction errors aggregate as the horizon becomes larger. We choose 20 for the horizon, which balances the prediction errors and action performance with short latency.

\begin{table}[t]
  \renewcommand\arraystretch{1.2}
  \small
  \caption{Execution Overhead}
%   \vspace{-0.1in}
  \centering
  \begin{tabular}{|c|c|c|c|}
    \hline
    & \shortstack{Model\\Inference} &  \shortstack{Action\\Selectction}   & \shortstack{Total\\Latency} \\
    \hline
    Rule-based Method& N/A & N/A & 0.44 ms\\
    \hline
    Model-based DRL&  71.76 ms& 21.41 ms & 93.17 ms \\
    \hline
    Model-free DRL & N/A & N/A & 2.54 ms  \\
    \hline
    \aliasAPP & 294.72 ms   & 38.46 ms & 333.18 ms\\
    \hline

  \end{tabular}
  \label{inference_overhead}
%   \vspace{-0.2in}
\end{table}

\subsubsection{Execution Overhead}
Table~\ref{inference_overhead} tabulates the execution latency of different HVAC control methods for one timestep. The latency of the model-based method and \aliasAPP includes the prediction latency of the building dynamics model and the action selection latency spent by the controller. The latency of the rule-based method and model-free method is much smaller than the model-based method and \aliasAPP since they don't have dynamics model and controller.
The latency of the \aliasAPP model and controller are 294.72ms and 38.46ms respectively, which are both higher than the existing model-based method. First, the ENN model needs more time to evaluate the prediction results and calculate the weights of models in the ensemble learning of building dynamics models. Second, the MPPI controller needs to calculate the noise-related weight and evaluate different action sequences. However, from the results in Table \ref{inference_overhead}, we can see that \aliasAPP can find the best control action within one second, which can generate an executable control action fast, especially considering the control cycle is normally set to 15 minutes.

\section{Discussion}

\textbf{Building Model Calibration.}
Currently, we are leveraging the existing five-zone building model in EnergyPlus to evaluate all the existing control methods. We have not done the calibration for this building model since we have no historical operation data of that building. The buildings implemented in EnergyPlus are based on first principles thermodynamical models, so we expect this model to be similar in performance to a real building. Moreover, it is reasonable to compare all the control methods based on the same building model implemented in EnergyPlus as ground truth. So, for the evaluation done in the paper, we believe this is a fair comparison to test the relative performance of different schemes for “a particular building”. If the proposed \aliasAPP was to be deployed in a real building, we would first need to learn the dynamics model from the existing historical data from a real building. Then we deploy the model in the real building for control. If we were to do simulations to test \aliasAPP before real deployment, we need to develop a calibrated EnergyPlus model that matches the target building \cite{zhang2018practical, ding2019octopus}.

\textbf{Occupancy and Weather Model.}
In \aliasAPP, we provide the ground-truth value of weather and occupancy for ENN dynamics model. \aliasAPP might be a bit more optimistic since we assume perfect prediction for the weather and occupancy. The errors in prediction may impact controller performance. However, we believe the performance will not significantly deviate from actual results considering model prediction errors. First is that the existing occupancy and weather prediction model~\cite{beltran2014optimal},~\cite{winkler2020office},~\cite{chictu2019building},~\cite{stamatescu2016data} show very small prediction error. Second is that MPPI controller outputs the optimal trajectory over the planning horizon. MPPI only takes the first optimal action and re-plans at the next time
step based on new observations. This efficiently avoids compounding model error over time.  

\section{Conclusions}\label{sec:conclusion}
This paper proposes \aliasAPP, a novel model-based DRL HVAC control system for multi-zone buildings. We develop a new building dynamics model as an ensemble of multiple environment-conditioned neural network models. 
We also adopt a model predictive path integral control method to perform HVAC control. We compare the performance of \aliasAPP with the rule-based, and state-of-the-art model-based and model-free DRL schemes. The results show that \aliasAPP can achieve 10.65\%, 8.23\% energy savings on the former and comparable performance with the later, while maintaining (and sometimes even improving) thermal comfort of occupants.  Perhaps more importantly, we can achieve this by significantly reducing the training set required by an order of magnitude ($10.52\times$ less).

% \section*{Acknowledgments}
% This should be a simple paragraph before the References to thank those individuals and institutions who have supported your work on this article.

% {\appendix[Proof of the Zonklar Equations]
% Use $\backslash${\tt{appendix}} if you have a single appendix:
% Do not use $\backslash${\tt{section}} anymore after $\backslash${\tt{appendix}}, only $\backslash${\tt{section*}}.
% If you have multiple appendixes use $\backslash${\tt{appendices}} then use $\backslash${\tt{section}} to start each appendix.
% You must declare a $\backslash${\tt{section}} before using any $\backslash${\tt{subsection}} or using $\backslash${\tt{label}} ($\backslash${\tt{appendices}} by itself
%  starts a section numbered zero.)}


\bibliographystyle{IEEEtran}
\bibliography{bare_jrnl}

\newpage

 

% \end{IEEEbiography}



\vfill

\end{document}